\documentclass[pra,aps,onecolumn,twoside,superscriptaddress,notitlepage]{revtex4-1}

\usepackage{amsmath,amsfonts,amssymb,color,epsfig,graphics,graphicx,latexsym,theorem,url,verbatim}

\DeclareSymbolFont{extraup}{U}{zavm}{m}{n}
\DeclareMathSymbol{\varheart}{\mathalpha}{extraup}{86}
\DeclareMathSymbol{\vardiamond}{\mathalpha}{extraup}{87}

\newtheorem{definition}{Definition}
\newtheorem{proposition}[definition]{Proposition}
\newtheorem{lemma}[definition]{Lemma}

\newtheorem{theorem}[definition]{Theorem}
\newtheorem{corollary}[definition]{Corollary}
\newtheorem{conjecture}[definition]{Conjecture}

\newtheorem{remark}[definition]{Remark}
\newtheorem{example}[definition]{Example}
\newtheorem{question}[definition]{Question}

\def\squareforqed{\hbox{\rlap{$\sqcap$}$\sqcup$}}
\def\qed{\ifmmode\squareforqed\else{\unskip\nobreak\hfil
\penalty50\hskip1em\null\nobreak\hfil\squareforqed
\parfillskip=0pt\finalhyphendemerits=0\endgraf}\fi}
\def\endenv{\ifmmode\;\else{\unskip\nobreak\hfil
\penalty50\hskip1em\null\nobreak\hfil\;
\parfillskip=0pt\finalhyphendemerits=0\endgraf}\fi}
\newenvironment{proof}{\noindent \textbf{{Proof.~} }}{\qed}
\def\Dbar{\leavevmode\lower.6ex\hbox to 0pt
{\hskip-.23ex\accent"16\hss}D}
\makeatletter
\def\url@leostyle{%
  \@ifundefined{selectfont}{\def\UrlFont{\sf}}{\def\UrlFont{\small\ttfamily}}}
\makeatother
\def\bcj{\begin{conjecture}}
\def\ecj{\end{conjecture}}
\def\bcr{\begin{corollary}}
\def\ecr{\end{corollary}}
\def\bd{\begin{definition}}
\def\ed{\end{definition}}
\def\bea{\begin{eqnarray}}
\def\eea{\end{eqnarray}}
\def\bem{\begin{enumerate}}
\def\eem{\end{enumerate}}
\def\bex{\begin{example}}
\def\eex{\end{example}}
\def\bim{\begin{itemize}}
\def\eim{\end{itemize}}
\def\bl{\begin{lemma}}
\def\el{\end{lemma}}
\def\bpf{\begin{proof}}
\def\epf{\end{proof}}
\def\bpp{\begin{proposition}}
\def\epp{\end{proposition}}
\def\bqu{\begin{question}}
\def\equ{\end{question}}
\def\br{\begin{remark}}
\def\er{\end{remark}}
\def\bt{\begin{theorem}}
\def\et{\end{theorem}}

\def\btb{\begin{tabular}}
\def\etb{\end{tabular}}

\newcommand{\nc}{\newcommand}

\def\b{\beta}

\def\n{\nu}

\def\r{\rho}

\def\o{\omega}

 \nc{\bA}{{\bf A}} \nc{\bB}{{\bf B}}
 \nc{\bC}{{\mathbb{C}}}
 \nc{\bD}{{\bf D}} \nc{\bE}{{\bf E}} \nc{\bF}{{\bf F}}
 \nc{\bG}{{\bf G}} \nc{\bH}{{\bf H}} \nc{\bI}{{\bf I}}
 \nc{\bJ}{{\bf J}} \nc{\bK}{{\bf K}} \nc{\bL}{{\bf L}}
 \nc{\bM}{{\bf M}} \nc{\bN}{{\bf N}} \nc{\bO}{{\bf O}}
 \nc{\bP}{{\bf P}} \nc{\bQ}{{\bf Q}} \nc{\bR}{{\bf R}}
 \nc{\bS}{{\bf S}} \nc{\bT}{{\bf T}} \nc{\bU}{{\bf U}}
 \nc{\bV}{{\bf V}} \nc{\bW}{{\bf W}} \nc{\bX}{{\bf X}}
 \nc{\bZ}{{\bf Z}}

\nc{\cA}{{\cal A}} \nc{\cB}{{\cal B}} \nc{\cC}{{\cal C}}
\nc{\cD}{{\cal D}} \nc{\cE}{{\cal E}} \nc{\cF}{{\cal F}}
\nc{\cG}{{\cal G}} \nc{\cH}{{\cal H}} \nc{\cI}{{\cal I}}
\nc{\cJ}{{\cal J}} \nc{\cK}{{\cal K}} \nc{\cL}{{\cal L}}
\nc{\cM}{{\cal M}} \nc{\cN}{{\cal N}} \nc{\cO}{{\cal O}}
\nc{\cP}{{\cal P}} \nc{\cQ}{{\cal Q}} \nc{\cR}{{\cal R}}
\nc{\cS}{{\cal S}} \nc{\cT}{{\cal T}} \nc{\cU}{{\cal U}}
\nc{\cV}{{\cal V}} \nc{\cW}{{\cal W}} \nc{\cX}{{\cal X}}
\nc{\cZ}{{\cal Z}}

\nc{\hA}{{\hat{A}}} \nc{\hB}{{\hat{B}}} \nc{\hC}{{\hat{C}}}
\nc{\hD}{{\hat{D}}} \nc{\hE}{{\hat{E}}} \nc{\hF}{{\hat{F}}}
\nc{\hG}{{\hat{G}}} \nc{\hH}{{\hat{H}}} \nc{\hI}{{\hat{I}}}
\nc{\hJ}{{\hat{J}}} \nc{\hK}{{\hat{K}}} \nc{\hL}{{\hat{L}}}
\nc{\hM}{{\hat{M}}} \nc{\hN}{{\hat{N}}} \nc{\hO}{{\hat{O}}}
\nc{\hP}{{\hat{P}}} \nc{\hR}{{\hat{R}}} \nc{\hS}{{\hat{S}}}
\nc{\hT}{{\hat{T}}} \nc{\hU}{{\hat{U}}} \nc{\hV}{{\hat{V}}}
\nc{\hW}{{\hat{W}}} \nc{\hX}{{\hat{X}}} \nc{\hZ}{{\hat{Z}}}

\nc{\hn}{{\hat{n}}}

\def\tr{\mathop{\rm Tr}}
\def\w{\mathop{\rm W}}

\def\op{\oplus}

\newcommand{\bra}[1]{\langle#1|}
\newcommand{\ket}[1]{|#1\rangle}

\newcommand{\opr}[2]{|#1\rangle \langle #2|}

\def\Dbar{\leavevmode\lower.6ex\hbox to 0pt
{\hskip-.23ex\accent"16\hss}D}

\usepackage{graphicx}

\usepackage[frame,line,arrow,matrix,tips]{xy}	
\CompilePrefix{xygui-}
\makeindex

\def\w{\ar@{-}[l]}
\def\W{\ar@{=}[l]}

\def\A#1{\save []="#1" \restore}

\def\op#1{*+[F]{\rule[-0.2ex]{0ex}{2.1ex}#1}}	
\def\b{*={\vardiamond}}
\def\f{*={\bar{\vardiamond}}}
\def\r{*={\bullet}}
\def\o{*={\oplus}}

\def\n{*-{}\w}

\def\>{\rangle}
\def\<{\langle}

\def\meter{*+[]{\put(-3,0){\includegraphics[scale=.5]{meter}}~~~~}%
		\ar@{-}[l]}

\def\q#1{*+{\rule[-0.2ex]{0ex}{2.1ex}|#1\>}}
\def\qv#1#2{*+{\rule[-0.2ex]{0ex}{2.1ex}|#1\>=|#2\>}}
	
\def\gspace#1{*+{\rule[-0.2ex]{0ex}{2.1ex}%
	\setbox\sbox=\hbox{$#1$}%
	\hspace*{\wd\sbox}}}

\begin{document}

\title{Generalized Graph States Based on Hadamard Matrices}

\author{Shawn X Cui}
\affiliation{Department of Mathematics, University of California, Santa Barbara,
CA 93106, USA}

\author{Nengkun Yu}
\affiliation{Institute for Quantum Computing, University of Waterloo,
  Waterloo, Ontario, Canada}%
\affiliation{Department of Mathematics \& Statistics, University of
  Guelph, Guelph, Ontario, Canada}%

\author{Bei Zeng}
\affiliation{Institute for Quantum Computing, University of Waterloo,
  Waterloo, Ontario, Canada}%
\affiliation{Department of Mathematics \& Statistics, University of
  Guelph, Guelph, Ontario, Canada}%
\affiliation{Canadian Institute for Advanced Research, Toronto,
  Ontario, Canada}%

\begin{abstract}
Graph states are widely used in quantum information theory, including
entanglement theory, quantum error correction, and one-way quantum computing.
Graph states have a nice structure related to a certain graph, which is
given by either a stabilizer group or an encoding circuit, both can be directly given
by the graph.  To generalize graph states, whose stabilizer groups are
abelian subgroups of the Pauli group,
one approach taken is to study non-abelian stabilizers. In this work, we propose to
generalize graph states based on the encoding circuit, which is completely
determined by the graph and a Hadamard matrix. We study the entanglement structures
of these generalized graph states, and show that they are all maximally mixed locally.
We also explore the relationship between
the equivalence of Hadamard matrices and local equivalence of the corresponding generalized graph states. This leads to a natural generalization of the Pauli $(X,Z)$ pairs, which characterizes the
local symmetries of these generalized graph states.
Our approach is also naturally generalized to construct graph quantum codes which are beyond
stabilizer codes.
\end{abstract}

\maketitle

\section{Introduction}

An undirected graph $G$ with $n$ vertices and the edge set $E(G)$ corresponds to a unique $n$-qubit quantum state $\ket{\psi_G}$, which is called the graph state (corresponding to the graph $G$). Graph states are extensively studied and widely used in quantum information theory, due to its nice entanglement structures~\cite{briegel2001persistent,hein2004multiparty}. Certain kind of graph states (e.g. the cluster states) can be used as resource states for measurement-based quantum computing~\cite{raussendorf2001one}. And it is known that any stabilizer state is in fact equivalent to some graph states via local Clifford operations~\cite{schlingemann2001stabilizer}.

Graph states are also building blocks for a wide class of quantum error-correcting codes. For instance, it is natural to choose the basis of a stabilizer code using stabilizer states. Furthermore, by including ancilla qubits for encoding, graphs can be used to represent stabilizer codes, called the graph codes~\cite{schlingemann2001quantum}, and any stabilizer code is local Clifford equivalent to some graph code~\cite{schlingemann2001stabilizer}. Going beyond the stabilizer codes, one can use graph states as basis for the so called codeword stabilized quantum codes~\cite{cross2007codeword}, with which good nonaddictive codes may be constructed.

Graph states are also of interests to many-body physics. They naturally appear as ground states of gapped local Hamiltonians, which are given by commuting local projectors~\cite{nielsen2004optical}. These states are relatively easy to analyze, and may exhibit interesting properties such as topological orders~\cite{raussendorf2005long} and symmetry-protected topological orders~\cite{else2012symmetry,skrovseth2009phase,doherty2009identifying,zeng2014topological}, which are beyond the traditional symmetry-breaking orders.

There are two equivalent ways to define $\ket{\psi_G}$. One is from a stabilizer formalism. That is, for each vertex $i$ in $G$, assign a stabilizer generator,
\begin{equation}
\label{eq:stabilizer}
g_i=X_i\prod_{j\in N(i)} Z_j
\end{equation}
where $X_i$ ($Z_j$) is the Pauli $X$ ($Z$) operator acting on the $i$th ($j$th) qubit, and $N(i)$ denotes the qubits that are neighbours of $i$ in graph $G$. Each $g_i$ has eigenvalues $1$ and $-1$, and the $g_i$s are mutually commuting. Therefore there exists a unique quantum state $\ket{\psi_G}$ satisfying $g_i\ket{\psi_G}=\ket{\psi_G}$, i.e. $\ket{\psi_G}$ is the stabilizer state
with the stabilizer group generated by $g_i$s.

The other way is given by a circuit $\mathcal{U}$ that generates $\ket{\psi_G}$ from the product state $\ket{0}^{\otimes n}$, i.e.
$\mathcal{U}\ket{0^{\otimes n}}=\ket{\psi_G}$, where
\begin{equation}
\label{eq:circuit}
\mathcal{U}=\frac{1}{2^{n/2}}\prod_{ij\in E(G)} C^{Z}_{ij} H^{\otimes n}.
\end{equation}
Here $E(G)$ is the edge set of the graph $G$. $H$ is the Hadamard matrix
\begin{equation}
H=
\begin{pmatrix}
1&1\\1&-1
\end{pmatrix},
\end{equation}
and $C^{Z}$ is the two-qubit controlled-Z gate which a diagonal matrix given by
\begin{equation}
C^{Z}=
\begin{pmatrix}
1&0 & 0 & 0\\
0&1 & 0 & 0\\
0&0 & 1 & 0\\
0&0 & 0 & -1\\
\end{pmatrix}.
\end{equation}

These two ways are equivalent definitions for $\ket{\psi_G}$ given that
\begin{equation}
\mathcal{U}Z_i\mathcal{U}^{\dag}=g_i,\ \forall\ i.
\end{equation}

Graph states has a natural generalization to the qudit case, based on the generalized Pauli operators $X_d,Z_d$ satisfying the commutation relations of a quantum plane~\cite{zhou2003quantum}
\begin{equation}
X_dZ_d=q_dZ_dX_d,
\end{equation}
where $X_d, Z_d$ are defined by the maps $X_d | i \rangle = |i-1 \; (\textrm{mod} \; d ) \rangle$, $Z_d | i \rangle = q_d^i | i \rangle, i = 0,1, \cdots, d-1$, and  $q_d=e^{2\pi i/d}$. Based on this, Eq.~\eqref{eq:stabilizer} naturally generalizes to
\begin{equation}
\label{eq:sym}
g_i=(X_d)^{\dag}_i\prod_{j\in N(i)} (Z_d)_j,
\end{equation}
and
$g_i$s are mutually commuting. The corresponding unique qudit graph state,
denoted by $\ket{\psi_{G,\mathcal{F}_d}}$, is then given by $g_i\ket{\psi_{G,\mathcal{F}_d}}=\ket{\psi_{G,\mathcal{F}_d}},\ \forall\ i$. Also, the circuit given in Eq.~\eqref{eq:circuit} has a natural generalization by replacing $H$ by the discrete Fourier transform
\begin{equation}
\label{eq:fourier}
\mathcal{F}_d=\sum_{i,j=0}^{d-1} q_d^{ij}\ket{i}\bra{j},
\end{equation}
and replacing $C^Z$ by its generalized version
\begin{equation}
\label{eq:CZd}
C^{Z_d}=\sum_{i,j=0}^{d-1} q_d^{ij}\ket{ij}\bra{ij},
\end{equation}
and naturally
\begin{equation}
\mathcal{U}_dZ_i\mathcal{U}_d^{\dag}=g_i,\ \forall\ i,
\end{equation}
for
\begin{equation}
\mathcal{U}_d=\frac{1}{d^{n/2}}\prod_{ij\in E(G)} C^{Z_d}_{ij} \mathcal{F}_d^{\otimes n}.
\end{equation}

Recently, there have been considerations to generalize the graph (stabilizer) states beyond the abelian group structure of the Pauli group. One approach is to generalize the stabilizer formalism, by allowing non-commuting stabilizers. This includes the monomial stabilizer states~\cite{van2011monomial} and the XS-stabilizer states~\cite{ni2014non}, which describes some well-known many-body quantum states, for instance the Affleck-Kennedy-Lieb-Tasaki states~\cite{affleck1988valence} and the twisted quantum double model states~\cite{hu2013twisted}. However, because these is no longer a simple relationship between the stabilizers and the circuits (as Eq.~\eqref{eq:circuit}), the corresponding `stabilizer states' with non-commutign stabilizers lack a clear graph structure. Another approach is to generalize the Pauli $X$ operators as a certain kind of group action corresponding to an non-abelian group, and together with a generalized controlled-NOT operation, the correspond generalized graph states can then be defined on bipartite graphs which are directed~\cite{brell2014generalized}.

In this work, we propose a generalization of graph states based on Hadamard matrices. On the one hand, this is a very natural generalization, by observing the information `encoded' in Eqs~\eqref{eq:fourier}\eqref{eq:CZd}. That is, in the circuit $\mathcal{U}_d$, if one uses a Hadamard matrix $H$ (to replace $\mathcal{F}_d$), then one may further replace $C^{Z_d}$ by some generalized controlled-$Z$ operation which is defined by the entries of $H$. In this sense, our generalization will be based on the circuit approach instead of the stabilizer approach. On the other hand, (complex) Hadamard matrices themselves are of great mathematical interests, which has already be connected to various areas of study in quantum information science~\cite{werner2001all,tadej2006concise,horadam2007hadamard}.

The advantage of our generalization is its simple description at the first place: given an undirected graph $G$ with $n$ vertices, and an $d\times d$ (symmetric) Hadamard matrix $H$, a unique generalized graph state $\ket{\psi_{G,H}}$ is then defined. We focus on basic properties of these graph states, especially their structures related to the properties of the graph $G$ and the Hadamard matrix $H$.

For basic entanglement properties of $\ket{\psi_{G,H}}$, we show that $\ket{\psi_{G,H}}$ has maximally entangled single particle states regardless of the choice of the graph $G$. And $\ket{\psi_{G,H}}$ has a tensor network representation with tensors directly given by the entries of $H$. If $H$ has a tensor product structure, then $\ket{\psi_{G,H}}$ also has a tensor product structure.

Since one of the most basic properties of Hadamard matrices are their equivalence~\cite{horadam2007hadamard},
we explore the relationship between the equivalence of Hadamard matrices and local equivalence of the corresponding graph states. Our main results along this line include the following.
\begin{itemize}
\item $\ket{\psi_{G,H}}$ may not be local unitary equivalent to $\ket{\psi_{G,\mathcal{F}_n}}$ for some $G$.
\item Certain equivalence of $H$ corresponds to the local unitary equivalence of $\ket{\psi_{G,H}}$.
\item For any bipartite graph $G$, equivalence of $H$ corresponds to the local unitary equivalence of $\ket{\psi_{G,H}}$.
\item Certain symmetry (automorphism) of $H$ corresponds to the
local symmetries (stabilizers) of $\ket{\psi_{G,H}}$.
\end{itemize}

Furthermore, we show that the generalization of the circuit $\mathcal{U}_d$ can also be used as an encoding circuit for quantum error-correcting codes, by adding a classical encoder. This leads to non-stabilizer codes, where the effects of some errors are easy to analyze, depending on the structure of $H$.

\section{The generalized graph states}

\begin{definition}
A complex Hadamard matrix $H$ is a $d\times d$ matrix which satisfies that each matrix element $h_{ij}$ of $H$ for $i,j=0,1,\ldots,d-1$ with $|h_{ij}|=1$, and
\begin{equation}
H^{\dag}H=dI_d,
\end{equation}
where $I_d$ is the $d\times d$ identity matrix.
\end{definition}

We consider a $d\times d$ complex Hadamard matrix $H$ that is symmetric, i.e.
\begin{equation}
H=H^T,\quad H^*H=dI_d.
\end{equation}
For any quantum state in $\mathbb{C}^d\otimes\mathbb{C}^d$, we define a generalized controlled-$Z$ gate, which is completely determined by $H$. For this reason we write this gate by $C^{H}$, which is given by
\begin{equation}
\label{eq:CH}
C^{H}\ket{ij}=h_{ij}\ket{ij}.
\end{equation}
The reason for choosing $H$ symmetric is that $C^{H}$ does not distinguish the controlled qudit from the target qubit, so one can then define a generalized graph state on an undirected graph, which is given by the following definition.

\begin{definition}
For an undirected graph $G$ of $n$ vertices, with vertex set $V(G)$ and
edge set $E(G)$. Define a circuit based on the symmetric Hadamard matrix $H$ by
\begin{equation}
\mathcal{U}_{G,H}=\frac{1}{d^{n/2}}\prod_{ij\in E(G)} C^{H}_{ij} H^{\otimes n},
\end{equation}
where $C^{H}_{ij}$ is the generalized controlled-$Z$ gate acting on the $i,j$th qudits,
and the $n$-qudit generalized graph state $\ket{\psi_{G,H}}$ given by
\begin{equation}
\label{eq:GS}
\ket{\psi_{G,H}}=\mathcal{U}_{G,H}\ket{0}^{\otimes n}.
\end{equation}
\end{definition}

For $d=2$ and
$
H=
\begin{pmatrix}
1&1\\1&-1
\end{pmatrix}$,
$\ket{\psi_{G,H}}$ is the usual qubit graph state $\ket{\psi_{G}}$.
And when $H$ is the $d$-point discrete Fourier transform $\mathcal{F}_n$,
$\ket{\psi_{G,H}}$ is the usual qudit graph state $\ket{\psi_{G,\mathcal{F}_n}}$.

We also introduce a graphical way to represent the circuit $\mathcal{U}_{G,H}$,
which will helps us to visualize/prove some general properties of $\ket{\psi_{G,H}}$.
Based on the usual way of drawing a quantum circuit, we further use $H$ representing the unitary
transform $\frac{1}{\sqrt{d}}H$, and the black-diamonds to replace the black-dots in the usual
controlled-$Z$, to represent the generalized controlled-$Z$, given by $C^H$. As an example,
for the triangle graph of Figure $1(c)$, we have the corresponding circuit for creating $\ket{\psi_{\triangle,H}}$
as given in Figure $2$.

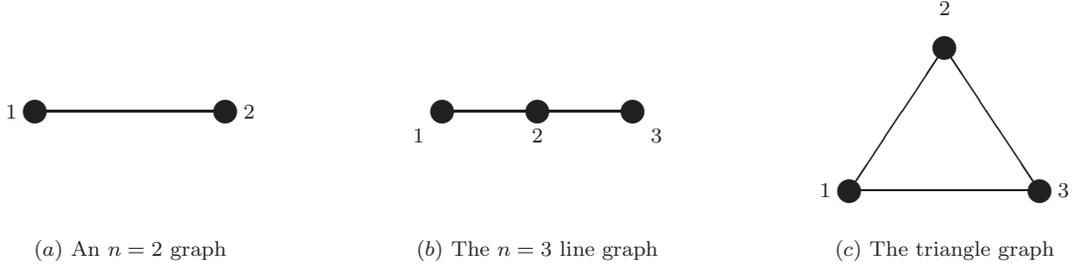
\begin{figure}[h!]
\centerline{
\footnotesize\unitlength3\unitlength%
\begin{tabular}{ccc}
\begin{picture}(50,35)(0,5)
\put(13,20){\circle*{3}}
\put(10,20){\makebox(0,0){$1$}}
\put(37,20){\circle*{3}}
\put(40,20){\makebox(0,0){$2$}}
\put(37,20){\line(-1,0){24}}
\put(13,20){\line(1,0){24}}
\end{picture}
&\unitlength1\unitlength
\begin{picture}(50,50)(0,5)
\put(25,20){\circle*{3}}
\put(25,17){\makebox(0,0){$2$}}
\put(13,20){\circle*{3}}
\put(10,17){\makebox(0,0){$1$}}
\put(37,20){\circle*{3}}
\put(40,17){\makebox(0,0){$3$}}
\put(37,20){\line(-1,0){24}}
\put(13,20){\line(1,0){24}}
\end{picture}
&\unitlength1\unitlength
\begin{picture}(50,50)(0,5)
\put(25,28){\circle*{3}}
\put(25,33){\makebox(0,0){$2$}}
\put(13,10){\circle*{3}}
\put(10,10){\makebox(0,0){$1$}}
\put(37,10){\circle*{3}}
\put(40,10){\makebox(0,0){$3$}}
\put(37,10){\line(-1,0){24}}
\put(13,10){\line(1,0){24}}
\put(13,10){\line(2,3){12}}
\put(25,28){\line(-2,-3){12}}
\put(37,10){\line(-2,3){12}}
\put(25,28){\line(2,-3){12}}
\end{picture}
\\
$(a)$ An $n=2$ graph
&
$(b)$ The $n=3$ line graph
&
$(c)$ The triangle graph
\end{tabular}}
\caption{Some $n=2$ and $n=3$ graphs}
\label{fig:fig1}
\end{figure}

\begin{figure}[h!]
\begin{center}

\def\gAxA{\op{H}\w\A{gAxA}}
\def\gAxB{\op{H}\w\A{gAxB}}
\def\gAxC{\op{H}\w\A{gAxC}}
\def\gBxA{\b\w\A{gBxA}}
\def\gBxB{\b\w\A{gBxB}}
\def\gCxB{\b\w\A{gCxB}}
\def\gCxC{\b\w\A{gCxC}}
\def\gDxA{\b\w\A{gDxA}}
\def\gDxC{\b\w\A{gDxC}}


\def\bA{{\ket{0}}}
\def\bB{{\ket{0}}}
\def\bC{{\ket{0}}}

~
\xymatrix@R=5pt@C=10pt{
    \bA & \gAxA &\gBxA &\n   &\gDxA &\n
\\  \bB & \gAxB &\gBxB &\gCxB &\n   &\n
\\  \bC & \gAxC &\n   &\gCxC &\gDxC &\n
%
%
\ar@{-}"gBxA";"gBxB"
\ar@{-}"gCxB";"gCxC"
\ar@{-}"gDxA";"gDxC"
}
\label{fig:triangleEn}
\caption{Circuit for creating a triangle graph state. $H$ represents the unitary
transform $\frac{1}{\sqrt{d}}H$, and the black-diamonds connected
by a line represents the generalized controlled-$Z$ gate $C^H$.}
\end{center}
\end{figure}
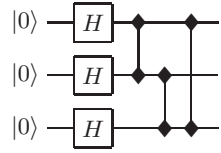

In order to discuss the properties of $\ket{\psi_{G,H}}$, we would need the concepts of local equivalence of two quantum states.

\begin{definition}
Two $n$-qudit quantum states $\ket{\psi_1}$ and $\ket{\psi_2}$ are local unitary (LU) equivalent if there exists a local unitary operator $U=\bigotimes_{i=1}^n U_i$, such that $U\ket{\psi_1}=\ket{\psi_2}$, where each $U_i$  is a single-qudit unitary operation.
\end{definition}

The single qudit Clifford group is the automorphism group of the qudit Pauli group generated by $X_d$ and $Z_d$.
\begin{definition}
Two $n$-qudit quantum states $\ket{\psi_1}$ and $\ket{\psi_2}$ are local Clifford (LC) equivalent if there exists a local unitary operator $L=\bigotimes_{i=1}^n L_i$, such that $L\ket{\psi_1}=\ket{\psi_2}$, where each $L_i$ is a single-qudit Clifford operation.
\end{definition}

\section{Entanglement properties}

We discuss basic entanglement properties of $\ket{\psi_{G,H}}$.
Denote $\Gamma_i$ the $d\times d$ diagonal matrix whose diagonal elements are
the $i$th column of $H$. Denote $\ket{\text{GHZ}_{n,d}}$ the $n$-qudit GHZ state, which is given by
\begin{equation}
\ket{\text{GHZ}_{n,d}}=\frac{1}{\sqrt{d}} \sum_{i=1}^d\ket{{\overbrace{ii\ldots i}^{n}}}.
\end{equation}

\subsection{The bipartite system}
We start to examine the properties of $\ket{\psi_{G,H}}$ for $n=2$, where
the nontrivial graph corresponds to the one given by Figure $1(a)$.
In this case, the corresponding
$\ket{\psi_{G,H}}$ is a maximally entangled state.
This can be seen from
\begin{eqnarray}
\ket{\psi_{G,H}}&=&\frac{1}{d}C^{H}_{12}
H^{\otimes 2}\ket{0}^{\otimes 2}=\frac{1}{d}C^{H}_{12}(\Gamma_1\otimes \Gamma_1)\sum_{ij}\ket{ij}
=(\Gamma_1\otimes \Gamma_1)\frac{1}{d}C^{H}_{12}\sum_{ij}\ket{ij}\nonumber\\
&=&(\Gamma_1\otimes \Gamma_1)\frac{1}{\sqrt{d}}\sum_{i}\ket{i}\left(\frac{1}{\sqrt{d}}\sum_j h_{ij}\ket{j}\right)=(\Gamma_1\otimes \Gamma_1)\frac{1}{\sqrt{d}}\sum_{i}\ket{i}\ket{\psi_i}.
\end{eqnarray}
Here the states $\ket{\psi_i}=H\ket{i}=\frac{1}{\sqrt{d}}\sum_j h_{ij}\ket{j}$ are orthonormal ($\bra{\psi_i}\psi_j\rangle=\delta_{ij}$) due to the
orthogonality of the rows of $H$.

It is obvious that $\ket{\psi_{G,H}}$ is LU equivalent to the state $\frac{1}{\sqrt{d}}\sum_{i}\ket{i}\ket{\psi_i}$, which is also a generalized graph state with the same graph but another Hadamard matrix with all $1$ elements of the first row/column. In other words, to discuss entanglement properties of  $\ket{\psi_{G,H}}$, it suffices to assume that $\Gamma_1=I$. In fact, this $\Gamma_1=I$ assumption is without loss of generality, as in general $\ket{\psi_{G,H}}$ will be LU equivalent another generalized graph state with the same graph whose Hadamard matrix is with all $1$ elements of the first row/column (see Lemma~\ref{lm:phase} and Theorem~\ref{th:Peq}). Therefore, from we will just assume $\Gamma_1=I$ for all the Hadamard matrix $H$ in our following discussions.

Furthermore, since we also have (for $\Gamma_1=I$)
\begin{eqnarray}
\ket{\psi_{G,H}}=\frac{1}{d}C^{H}_{12}\sum_{ij}\ket{ij}
=
\frac{1}{\sqrt{d}}\left(\frac{1}{\sqrt{d}}\sum_{j}\sum_j H_{ij}\ket{i}\right)\ket{j}=\frac{1}{\sqrt{d}}\sum_{j}\ket{\psi_j}\ket{j},
\end{eqnarray}
we have that $H\otimes H^*$
(or $H^*\otimes H$) is a symmetry of $\ket{\psi_{G,H}}$. That is,
\begin{equation}
H\otimes H^*\ket{\psi_{G,H}}=\ket{\psi_{G,H}}.
\end{equation}

This shows that $\ket{\psi_{G,H}}$ is a maximally entangled state,
which is independent of the choice of $H$. Or in other world, all the $\ket{\psi_{G,H}}$ are
local unitary equivalent to each other. This is consistent with the observation
that all $\ket{\psi_{G,H}}$ can be used in teleportation and dense-coding schemes~\cite{werner2001all}.

\subsection{Single particle entanglement}

For $n=3$, there are two kinds of connected graphs $G$. The first one has edge set $E(G)=\{(12),(23)\}$,
given by the line graph of Figure $1(b)$. The other one has edge set $E(G)=\{(12),(23),(13)\}$,
given by the triangle graph of Figure $1(c)$. We discuss the line graph here and will discuss the triangle graph in
Sec. 5.

For $G$ being a three-qudit line graph, we have
\begin{equation}
\ket{\psi_{G,H}}=\frac{1}{d^{3/2}}C^{H}_{12}C^{H}_{23}\sum_{ijk}\ket{ijk}
=\frac{1}{\sqrt{d}}\sum_{j}\ket{\psi_j}\ket{j}\ket{\psi_j},
\end{equation}
which is LU equivalent to $\ket{\text{GHZ}_{3,d}}$. This is also independent of
the choice of $H$, i.e. all these $\ket{\psi_{G,H}}$ are
LU equivalent to each other.

This property generalizes to multi-qudit case, which is given by the following proposition.
\begin{proposition}
\label{pro:star}
If $G$ is an $n$-qudit star-shape graph, i.e. with edge set $E(G)=\{(12),(13),\ldots,(1n)\}$, then $\ket{\psi_{G,H}}$ is LU equivalent to $\ket{\text{GHZ}_{n,d}}$.
\end{proposition}

\textbf{Proof}: Notice that
\begin{eqnarray}
\ket{\psi_{G,H}}&=&\frac{1}{d^{n/2}}\prod_{ij\in E(G)} C^{H}_{ij} H^{\otimes n}\ket{0}^{\otimes n}\nonumber\\
&=&\frac{1}{d^{n/2}}\prod_{j=2}^{n} C_{1j}^H\sum\ket{{i_1i_2\ldots i_n}}
\nonumber\\
&=&\frac{1}{\sqrt{d}}\sum_{j}\ket{j}\ket{\psi_j}\cdots\ket{\psi_j},
\end{eqnarray}
which is LU equivalent to $\ket{\text{GHZ}_{n,d}}$ $\square$.

A direct consequence of Proposition~\ref{pro:star} is the following
\begin{corollary}
Any single particle reduced density matrix of
 $\ket{\psi_{G,H}}$ is maximally mixed for any connected graph $G$.
\end{corollary}

\textbf{Proof}: Denote the vertex set of the graph $G$ by $V(G)$. For any vertex $a\in V(G)$, denote $G_a^{\star}$ the graph with the same vertices
as that of $G$, but only edges $aj\in E(G)$. Without loss of generality we only consider the case of $a=1$. Then we have
\begin{eqnarray}
\ket{\psi_{G,H}}&=& \prod_{i\neq 1,\ ij\in V(G)}C_{ij}^{H}\ket{\psi_{G_1^{\star},H}}.
\end{eqnarray}
Since $G$ is a connected graph, according to Proposition~\ref{pro:star}, $\ket{\psi_{G_1^{\star},H}}$ is LU equivalent to a tensor product of some $\ket{\text{GHZ}_{m,d}}$s (for $m\leq n$). Furthermore, $\prod_{i\neq 1,\ ij\in V(G)}C_{ij}^{H}$ does not act on the $1$st qudit. Consequently, the single particle reduced density matrix of the $1$st qudit is then maximally mixed $\square$.

\subsection{The tensor network representation}

It is known that the graph states are `finitely correlated states'~\cite{fannes1992finitely,verstraete2004valence} with a tensor network representation~\cite{perez2008peps}. They are unique ground states of Hamiltonian of local commuting projectors with locality determined by the graph $G$. Here we show that these properties naturally carry over to the generalized graph states $\ket{\psi_{G,H}}$.

First of all, it is straightforward to show that $\ket{\psi_{G,H}}$ is a unique ground state of gapped Hamiltonian of commuting projectors. This is because that we know $\ket{0}^{\otimes n}$ is stabilized by $\{\ket{0_i}\bra{0_i}\}_{i=1}^n$, where $\ket{0_i}$ is $\ket{0}$ state of the $i$th qudit. Since $\ket{\psi_{G,H}}=\mathcal{U}_{G,H}\ket{0}^{\otimes n}$, $\ket{\psi_{G,H}}$ is then stabilized by $\{\mathcal{U}_{G,H}\ket{0_i}\bra{0_i}\mathcal{U}_{G,H}^{\dag}\}_{i=1}^n$. Therefore, $\ket{\psi_{G,H}}$ is the unique ground state of the gapped Hamiltonian
\begin{equation}
\mathfrak{H}=-\sum_{i=1}^n \mathcal{U}_{G,H}\ket{0_i}\bra{0_i}\mathcal{U}_{G,H}^{\dag},
\end{equation}
where each term $\mathcal{U}_{G,H}\ket{0_i}\bra{0_i}\mathcal{U}_{G,H}^{\dag}$ are mutually commuting. Furthermore, the locality of each $\mathcal{U}_{G,H}\ket{0_i}\bra{0_i}\mathcal{U}_{G,H}^{\dag}$ is determined by the connectivity of the graph $G$, given the structure of $\mathcal{U}_{G,H}$.

$\ket{\psi_{G,H}}$ has a representation as a tensor product state (also called the projective entanglement-pair states (PEPS)). To discuss this representation, we first choose the (un-normalized) `bond state' between the sites $s,t$ to be
\begin{equation}
\ket{\psi^{\text{bond}}_{st}}=C_{st}^{H}\sum_{i_si_t}\ket{i_si_t},
\end{equation}
where $i_s,i_t\in (0,1,\ldots,d-1)$.

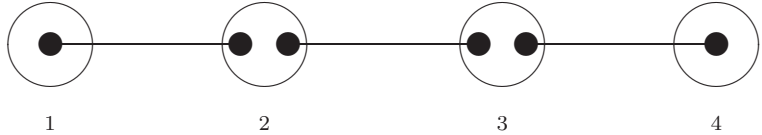
\begin{figure}[h!]
\centerline{
\footnotesize\unitlength3\unitlength%
\begin{picture}(50,50)(0,5)
\put(-17,20){\circle{10}}
\put(-17,20){\circle*{3}}
\put(-17,10){\makebox(0,0){$1$}}
\put(7,20){\circle*{3}}
\put(10,20){\circle{10}}
\put(10,10){\makebox(0,0){$2$}}
\put(7,20){\line(-1,0){24}}
\put(-17,20){\line(1,0){24}}
\put(13,20){\circle*{3}}
\put(37,20){\circle*{3}}
\put(40,20){\circle{10}}
\put(40,10){\makebox(0,0){$3$}}
\put(37,20){\line(-1,0){24}}
\put(13,20){\line(1,0){24}}
\put(43,20){\circle*{3}}
\put(67,10){\makebox(0,0){$4$}}
\put(67,20){\circle*{3}}
\put(67,20){\circle{10}}
\put(67,20){\line(-1,0){24}}
\put(43,20){\line(1,0){24}}
\end{picture}}
\caption{$\ket{\Psi_G}$ for a graph $G$ of four vertices and $E(G)=(12,23,34)$. Each
circle represent a site. Each black dot represent a qudit. And two black dots connected by a line
represent a bond $\ket{\psi^{\text{bond}}}$.}
\label{fig:PEPS}
\end{figure}

Consider a graph $G$. For each site $s\in V(G)$, denote $m(s)$ the degree of the vertex $s$ in $G$. Now consider a state $\ket{\Psi_G}$ which has $m(s)$ qudits on the site $s$, given by
\begin{equation}
\ket{\Psi_G}=\bigotimes_{st\in E(G)}\ket{\psi_{st}^{\text{bond}}}.
\end{equation}
As an example, $\ket{\Psi_G}$ for a graph $G$ of four vertices and $E(G)=(12,23,34)$ is illustrated in Fig.~\ref{fig:PEPS}.
\vspace{-2.5cm}

\begin{proposition} $\ket{\psi_{G,H}}$ has the following representation
\begin{equation}
\ket{\psi_{G,H}}\propto\prod_{s\in V(G)}\left(\sum_{i_s=1}^d\ket{i_s}\bra{\overbrace{i_si_s\dots i_s}^{m(s)}}\right)\ket{\Psi_G}.
\end{equation}
\end{proposition}

\textbf{Proof}: Notice that
\begin{eqnarray}
\ket{\psi_{G,H}}&\propto&\prod_{st\in E(G)}C^H_{st}H^{\otimes n}\ket{0}^{\otimes n}\nonumber\\
&=& \prod_{st\in E(G)}C^H_{st}\sum_{i_1i_2\ldots i_n}\ket{i_1i_2\ldots i_n}.
\end{eqnarray}
On the other hand,
\begin{eqnarray}
&&\prod_{s\in V(G)}\left(\sum_{i_s=1}^d\ket{i_s}\bra{\overbrace{i_si_s\dots i_s}^{m(s)}}\right)\ket{\Psi_G}\nonumber\\
&=&\prod_{s\in V(G)}\left(\sum_{i_s=1}^d\ket{i_s}\bra{\overbrace{i_si_s\dots i_s}^{m(s)}}\right)
\left(\bigotimes_{rt\in E(G)}
C_{rt}^{H}\sum_{i_ri_t}\ket{i_ri_t}\right)\nonumber\\
&=& \prod_{st\in E(G)}C^H_{st}\sum_{i_1i_2\ldots i_n}\ket{i_1i_2\ldots i_n}.
\end{eqnarray}

We remark that this tensor network representation may help to analyze what kind of generalized graph states may be resource states for measurement-based quantum computing~\cite{gross2007novel}.

\subsection{The tensor product structure}

It is easy to show that if $H_1$ and $H_2$ are Hadamard matrices, then $H=H_1\otimes H_2$ is also a Hadamard matrix. A natural question is then what is the relationship between the structure of $\ket{\psi_{G,H}}$ and those of $\ket{\psi_{G,H_1}}$ and $\ket{\psi_{G,H_2}}$. This is given by the following proposition.

\begin{proposition}
\label{pro:tensor}
If $H=H_1\otimes H_2$, then $\ket{\psi_{G,H}}=\ket{\psi_{G,H_1}}\otimes \ket{\psi_{G,H_2}}$ (up to qudit permutation).
\end{proposition}

\textbf{Proof}: Let the dimensions of $H_1, H_2$ be $d_1, d_2$, respectively. Since $H_1$ and $H_2$ are both Hadamard matrices, then $H=H_1\otimes H_2$ is also a Hadamard matrix of dimension $d_1d_2$. We identify the Hilbert space $\bC^{d_1d_2}$ with $\bC^{d_1} \otimes \bC^{d_2}$. Then $H^{\otimes n}\ket{0}^{\otimes n}$ can be naturally interpreted as $(H_1\otimes H_2)^{\otimes n}\ket{00}^{\otimes n}$. We then need to examine the generalized controlled-$Z$, which reads
\begin{equation}
C^{H}\ket{i_1i_2,j_1j_2}
=H_{i_1i_2,j_1j_2}\ket{i_1i_2,j_1j_2}
=(H_1)_{i_1j_1}\ket{i_1j_1}\otimes (H_2)_{i_2j_2}\ket{i_2j_2}
= C^{H_1}\ket{i_1j_1}\otimes C^{H_2}\ket{i_2j_2}
\end{equation}

Then it is clear that under this identification of qudits, we have $\ket{\psi_{G,H}}=\ket{\psi_{G,H_1}}\otimes \ket{\psi_{G,H_2}}$ $\square$.

\begin{figure}[h!]

\def\gAxA{\op{H}\w\A{gAxA}}
\def\gAxB{\op{H}\w\A{gAxB}}
\def\gAxC{\op{H}\w\A{gAxC}}
\def\gAxD{\op{H}\w\A{gAxD}}
\def\gAxE{\op{H}\w\A{gAxE}}
\def\gAxF{\op{H}\w\A{gAxF}}
\def\gBxA{\r\w\A{gBxA}}
\def\gBxC{\r\w\A{gBxC}}
\def\gBxB{*-{}\w\A{gBxB}}
\def\gBxD{*-{}\w\A{gBxD}}
\def\gCxB{\r\w\A{gCxB}}
\def\gCxD{\r\w\A{gCxD}}
\def\gCxC{*-{}\w\A{gCxC}}
\def\gBxE{*-{}\w\A{gBxE}}
\def\gDxC{\r\w\A{gDxC}}
\def\gDxE{\r\w\A{gDxE}}
\def\gDxD{*-{}\w\A{gDxD}}
\def\gBxF{*-{}\w\A{gBxF}}
\def\gExD{\r\w\A{gExD}}
\def\gExF{\r\w\A{gExF}}
\def\gDxB{*-{}\w\A{gDxB}}
\def\gFxF{*-{}\w\A{gFxF}}
\def\gCxA{*-{}\w\A{gCxA}}
\def\gDxA{*-{}\w\A{gDxA}}
\def\gExA{*-{}\w\A{gExA}}
\def\gFxA{*-{}\w\A{gFxA}}
\def\gGxA{\r\w\A{gGxA}}
\def\gGxE{\r\w\A{gGxE}}
\def\gHxE{*-{}\w\A{gHxE}}
\def\gGxF{*-{}\w\A{gGxF}}
\def\gHxB{\r\w\A{gHxB}}
\def\gHxF{\r\w\A{gHxF}}

\def\bA{|0\rangle_{1_a}}
\def\bB{|0\rangle_{1_b}}
\def\bC{|0\rangle_{2_a}}
\def\bD{|0\rangle_{2_b}}
\def\bE{|0\rangle_{3_a}}
\def\bF{|0\rangle_{3_b}}

~~~~~
\xymatrix@R=5pt@C=10pt{
    \bA & \gAxA &\gBxA &\gCxA &\gDxA &\gExA &\gFxA &\gGxA &\n   &\n
\\  \bB & \gAxB &\gBxB &\gCxB &\gDxB &\n   &\n   &\n   &\gHxB &\n
\\  \bC & \gAxC &\gBxC &\gCxC &\gDxC &\n   &\n   &\n   &\n   &\n
\\  \bD & \gAxD &\gBxD &\gCxD &\gDxD &\gExD &\n   &\n   &\n   &\n
\\  \bE & \gAxE &\gBxE &\n   &\gDxE &\n   &\n   &\gGxE &\gHxE &\n
\\  \bF & \gAxF &\gBxF &\n   &\n   &\gExF &\gFxF &\gGxF &\gHxF &\n
\ar@{-}"gBxA";"gBxC"
\ar@{-}"gCxB";"gCxD"
\ar@{-}"gDxC";"gDxE"
\ar@{-}"gExD";"gExF"
\ar@{-}"gGxA";"gGxE"
\ar@{-}"gHxB";"gHxF"
}
\caption{The circuit for generating the generalized graph state $\ket{\psi_{\triangle,H'}}$
with the Hadamard matrix given in Eq.~\eqref{eq:hadamarT}. $\ket{i_a}\ket{i_b}$
for $i=1,2,3$ is the input $\ket{0}$ state for the $i$th qudit.}
\label{fig:case3}
\end{figure}
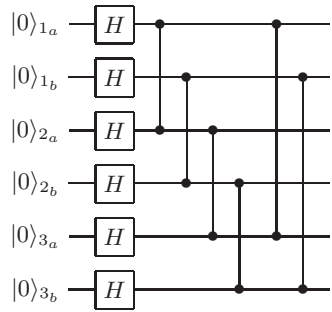

As an example, consider a triangle graph with the Hadamard matrix $H'$ given by
\begin{equation}
\label{eq:hadamarT}
\begin{pmatrix}
1&1\\1&-1
\end{pmatrix}\otimes \begin{pmatrix}
1&1\\1&-1
\end{pmatrix}.
\end{equation}
The circuit for generating the corresponding generalized graph state $\ket{\psi_{\triangle,H'}}$ is then given in
Fig.~\ref{fig:case3}, where $H$ represents the single-qubit
unitary operation
$\frac{1}{\sqrt{2}}\begin{pmatrix}
1&1\\1&-1
\end{pmatrix}$, and the black dots connected by a line is just the usual controlled-$Z$ gate of two qubits.

\section{Local equivalence and symmetry}

An important basic property of Hadamard matrices is their equivalence.
\begin{definition}
Two $d\times d$ Hadamard matrices $H_1$  and $H_2$ are equivalent if there exists
two $d\times d$ permutation matrices $P_1,P_2$, and two diagonal
matrices $D_1,D_2$, such that
\begin{equation}
H_1=D_1P_1H_2P_2D_2.
\end{equation}
\end{definition}

The classification of complex Hadamard matrices for $d=2,3,4,5$ up to equivalence
is given by the following theorem (see, e.g.~\cite{horadam2007hadamard}).
\begin{theorem}
For $d=2,3,5$, any complex Hadamard matrix are equivalent to the discrete Fourier
transform $\mathcal{F}_d$. For $d=4$, any complex Hadamard matrix
is equivalent to $H_{\alpha}$ given by
\begin{equation}
\label{eq:hadamardd4}
H_{\alpha}=
\begin{pmatrix}
1&1&1&1\\ 1&1&-1&-1\\1&-1&e^{i\alpha}&-e^{i\alpha}\\1&-1&-e^{i\alpha}&e^{i\alpha}
\end{pmatrix},
\end{equation}
where $\alpha\in\mathbb{R}$.
\end{theorem}

\subsection{The local equivalence}

We now study the relationship between the equivalence of Hadamard matrices and the
LU equivalence of the corresponding generalized graph states. We first look at the
relationship between the corresponding generalized controlled-$Z$ operations $C^H$.
This will be given by the following two lemmas.
\begin{lemma}
\label{lm:phase}
For two equivalent $d\times d$ Hadamard matrices $H_2=D_1H_1D_2$, where $D_1,D_2$ are $d\times d$ diagonal matrices,
\begin{equation}
C^{H_2}=(D_1\otimes D_2)C^{H_1}.
\end{equation}
\end{lemma}
\textbf{Proof}: Assume $D_1=\sum_{i=0}^{d-1}d_{1,i}\opr{i}{i}$ and $D_2=\sum_{i=0}^{d-1}d_{2,i}\opr{i}{i}$.
And denote $h_{ij}^{(1)}$ ($h_{ij}^{(2)}$) the matrix elements of $C^{H_1}$ ($C^{H_2}$). Then we have
\begin{equation}
C^{H_2}\ket{ij}=\sum_{i,j=0}^{d-1}h_{ij}^{(2)}\ket{ij}=\sum_{i,j=0}^{d-1}d_{1,i}d_{2,j}h_{ij}^{(1)}\ket{ij}=(D_1\otimes D_2)C^{H_1}\ket{ij}.\quad \square
\end{equation}
We illustrate this relationship between $C^{H_1}$ and $C^{H_2}$ in Fig.~\ref{fig:CH1}.

\begin{figure}[h!]
\begin{center}

\def\gAxA{\b\w\A{gAxA}}
\def\gAxB{\b\w\A{gAxB}}
\def\gBxA{*-{}\w\A{gBxA}}
\def\gBxB{*-{}\w\A{gBxB}}
\def\gCxA{\b\w\A{gCxA}}
\def\gCxB{\b\w\A{gCxB}}
\def\gDxA{\op{D_1}\w\A{gDxA}}
\def\gDxB{\op{D_2}\w\A{gDxB}}

\def\bA{\qv{q_{1}}{0}}
\def\bB{\qv{q_{2}}{0}}

~~~~~~~~~\xymatrix@R=5pt@C=10pt{
     & \gAxA &\gBxA\gBxA&~~~~&\gCxA &\gDxA &\n
\\   &  & &\longrightarrow~~& & & &
\\   & \gAxB &\gBxA\gBxA&~~~~&\gCxB &\gDxB &\n
\ar@{-}"gAxA";"gAxB"
\ar@{-}"gCxA";"gCxB"
}
\caption{A graphical way for illustrating the relationship between $C^{H_1}$ and $C^{H_2}$
for $H_2=D_1H_1D_2$. The left side represents $C^{H_1}$, and the right side represent
$C^{H_2}$ in terms of $C^{H_1}$.}
\label{fig:CH1}
\end{center}
\end{figure}

\begin{lemma}
\label{lm:P}
For two equivalent $d\times d$ Hadamard matrices $H_2=P_1H_1P^T_2$, where $P_1,P_2$ are $d\times d$ permutation matrices,
\begin{equation}
C^{H_2}=(P_1\otimes P_2)C^{H_1}(P_1^{T}\otimes P_2^T).
\end{equation}
\end{lemma}
\textbf{Proof}: Denote $\tilde{i}=P_1i$ and $\tilde{j}=P_2j$.
Notice that $H_2=\sum_{i,j=0}^{d-1}h_{i,j}^{(2)}\opr{i}{j}=\sum_{i,j=0}^{d-1}h_{i,j}^{(1)}P_1\opr{i}{j}P_2$, then $h_{ij}^{(2)}=h_{\tilde{i}\tilde{j}}^{(1)}$. Therefore we have
\begin{eqnarray}
C^{H_2}&=&\sum_{i,j=0}^{d-1}h_{i,j}^{(2)}\opr{ij}{ij}
=\sum_{i,j=0}^{d-1}h_{\tilde{i}\tilde{j}}^{(1)}\opr{ij}{ij}\nonumber\\
&=&((P_1\otimes P_2))\sum_{i,j=0}^{d-1}h_{\tilde{i}\tilde{j}}^{(1)}(P_1^T\otimes P_2^T)\opr{ij}{ij}(P_1\otimes P_2)(P_1^T\otimes P_2^T)\nonumber\\
&=&(P_1\otimes P_2)C^{H_1}(P_1^T\otimes P_2^T).\quad\square
\end{eqnarray}
We illustrate this relationship between $C^{H_1}$ and $C^{H_2}$ in Fig.~\ref{fig:CH2}.
\begin{figure}[h!]

\def\gAxA{\b\w\A{gAxA}}
\def\gAxB{\b\w\A{gAxB}}
\def\gBxA{\op{P_1}\w\A{gBxA}}
\def\gBxB{\op{P_2}\w\A{gBxB}}
\def\gCxA{\b\w\A{gCxA}}
\def\gCxB{\b\w\A{gCxB}}
\def\gDxA{\op{P_1^T}\w\A{gDxA}}
\def\gDxB{\op{P_2^T}\w\A{gDxB}}

\def\bA{\qv{q_{1}}{0}}
\def\bB{\qv{q_{2}}{0}}

~~~~~~~~~\xymatrix@R=5pt@C=10pt{
     & \gAxA&\n&~~~~&\gBxA &\gCxA &\gDxA &\n
\\   &  & & \longrightarrow~~& & & &
\\   & \gAxB&\n&~~~~&\gBxB &\gCxB &\gDxB &\n
\ar@{-}"gAxA";"gAxB"
\ar@{-}"gCxA";"gCxB"
}
\caption{A graphical way for illustrating the relationship between $C^{H_1}$ and $C^{H_2}$
for $H_2=P_1H_1P_2$. The left side represents $C^{H_1}$, and the right side represent
$C^{H_2}$ in terms of $C^{H_1}$.}
\label{fig:CH2}
\end{figure}

To study the relationship between the LU equivalence of the corresponding generalized graph
states $\ket{\psi_{G,H_1}}$ and $\ket{\psi_{G,H_2}}$, we will need the following concept
of $P$-equivalent Hadamard matrices.

\begin{definition}
Two Hadamard matrices $H_1$ and $H_2$ are called $P$-equivalent if there is a
$d\times d$ permutation $P$ and two $d\times d$ diagonal unitary matrices $D_1$
and $D_2$ such that
\begin{equation}
H_1=PD_1H_2D_2P^{T}.
\end{equation}
\end{definition}
Two $P$-equivalent Hadamard matrices are also equivalent,
but two equivalent Hadamard matrices are in general not $P$-equivalent.

We remark that the latter
is also true for two $P$-equivalent symmetric Hadamard matrices. That is,
two equivalent symmetric Hadamard matrices may not be $P$-equivalent.
A simple example is for $d=3$, and choose
\begin{equation}
\label{eq:Hqutrit}
H_1=\begin{pmatrix}
1 & 1 & 1\\
1 & \omega & \omega^2\\
1 & \omega^2 & \omega
\end{pmatrix}\quad\text{and}\quad
H_2=\begin{pmatrix}
1 & 1 & 1\\
1 & \omega^2 & \omega\\
1 & \omega & \omega^2
\end{pmatrix}=
\begin{pmatrix}
1 & 0 & 0\\
0 & 0 & 1\\
0 & 1 & 0
\end{pmatrix}H_1=PH_1,
\end{equation}
where $\omega=e^{2i\pi/3}$.
And obviously $H_1$ and $H_2$ are not $P$-equivalent.

We now show that, two $P$-equivalent symmetric Hadamard matrices correspond
to LU equivalent generalized graph states, for any graph $G$.

\begin{theorem}
\label{th:Peq}
If two symmetric Hadamard matrices $H_1$ and $H_2$ are $P$-equivalent, then $\ket{\psi_{G,H_1}}$ and $\ket{\psi_{G,H_2}}$ are LU equivalent.
\end{theorem}

\textbf{Proof}: Observe that for $H_1=PH_2P^{T}$, the corresponding $C^{H_1}$ and $C^{H_2}$ satisfy
$C^{H_1}=(P\otimes P)C^{H_2}(P\otimes P)^{T}$, as given by Lemma~\ref{lm:P}. Now notice that $PP^{T}=I$ and $(\sum_{i}\ket{i})^{\otimes n}$ is invariant under the action of $P$ on any qudit. Therefore, for Hadamard matrices $H_1,H2$ with $H_1=PH_2P^{T}$, $\ket{\psi_{G,H_1}}$ and $\ket{\psi_{G,H_2}}$ are LU equivalent.

For $H_1=D_1H_2D_2$ the corresponding $C^{H_1}$ and $C^{H_2}$ satisfy
$C^{H_1}=(D_1\otimes D_2)C^{H_2}$, as given by Lemma~\ref{lm:phase}. Therefore, for Hadamard matrices $H_1,H2$ with $H_1=D_1H_2D_2$, $\ket{\psi_{G,H_1}}$ and $\ket{\psi_{G,H_2}}$ are LU equivalent $\square$.

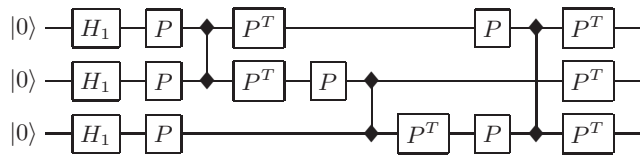
\begin{figure}[h!]

\def\gAxA{\op{H_1}\w\A{gAxA}}
\def\gAxB{\op{H_1}\w\A{gAxB}}
\def\gAxC{\op{H_1}\w\A{gAxC}}
\def\gBxA{\op{P}\w\A{gBxA}}
\def\gBxB{\op{P}\w\A{gBxB}}
\def\gBxC{\op{P}\w\A{gBxC}}
\def\gCxA{\b\w\A{gCxA}}
\def\gCxB{\b\w\A{gCxB}}
\def\gDxA{\op{P^T}\w\A{gDxA}}
\def\gDxB{\op{P^T}\w\A{gDxB}}
\def\gExB{\op{P}\w\A{gExB}}
\def\gFxB{\b\w\A{gFxB}}
\def\gFxC{\b\w\A{gFxC}}
\def\gGxC{\op{P^T}\w\A{gGxC}}
\def\gExA{*-{}\w\A{gExA}}
\def\gFxA{*-{}\w\A{gFxA}}
\def\gGxA{*-{}\w\A{gGxA}}
\def\gHxA{\op{P}\w\A{gHxA}}
\def\gHxC{\op{P}\w\A{gHxC}}
\def\gIxA{\b\w\A{gIxA}}
\def\gIxC{\b\w\A{gIxC}}
\def\gJxA{\op{P^T}\w\A{gJxA}}
\def\gGxB{*-{}\w\A{gGxB}}
\def\gHxB{*-{}\w\A{gHxB}}
\def\gIxB{*-{}\w\A{gIxB}}
\def\gJxB{\op{P^T}\w\A{gJxB}}
\def\gJxC{\op{P^T}\w\A{gJxC}}

\def\bA{{|0\rangle}}
\def\bB{{|0\rangle}}
\def\bC{{|0\rangle}}

~~~\xymatrix@R=5pt@C=10pt{
    \bA & \gAxA &\gBxA &\gCxA &\gDxA &\gExA &\gFxA &\gGxA &\gHxA &\gIxA &\gJxA &\n
\\  \bB & \gAxB &\gBxB &\gCxB &\gDxB &\gExB &\gFxB &\gGxB &\gHxB &\gIxB &\gJxB &\n
\\  \bC & \gAxC &\gBxC &\n   &\n   &\n   &\gFxC &\gGxC &\gHxC &\gIxC &\gJxC &\n
\ar@{-}"gCxA";"gCxB"
\ar@{-}"gFxB";"gFxC"
\ar@{-}"gIxA";"gIxC"
}
\caption{The LU equivalence of two generalized graph states $\ket{\psi_{\triangle,H_1}}$ and $\ket{\psi_{\triangle,H_2}}$. The two Hadamard matrices $H_1$ and $H_2$ satisfy $H_2=PH_1P^T$ for some permeation matrix $P$. Two black diamonds connected by a line represents $C^{H_1}$. This circuit generates the state $\ket{\psi_{\triangle,H_2}}$, which is given by Lemmas~\ref{lm:phase} and \ref{lm:P},
from the circuit generating  of $\ket{\psi_{\triangle,H_1}}$. Notice that each $PP^T=I$, so they do cancel. And $PH_1\ket{0}=DH_1\ket{0}$ for some diagonal matrix $D$, which commutes with all the diagonal $C^{H_1}$s. This then shows that $\ket{\psi_{\triangle,H_2}}=(P^TD)^{\otimes 3}\ket{\psi_{\triangle,H_1}}$.}
\label{fig:eqP}
\end{figure}

As an example, we illustrate the LU equivalence of two generalized graph states $\ket{\psi_{\triangle,H_1}}$ and $\ket{\psi_{\triangle,H_2}}$ for the triangle graph
in Fig.~\ref{fig:eqP}, where
the two Hadamard matrices $H_1$ and $H_2$ satisfy $H_2=PH_1P^T$ for some permeation matrix $P$.

A natural question is whether there exists a graph $G$, such that $\ket{\psi_{G,H_1}}$ and $\ket{\psi_{G,H_2}}$ are not LU equivalent for two equivalent symmetric Hadamard matrices $H_1$ and $H_2$. This is indeed possible. First of all, we only need to discuss the case that two symmetric Hadamard matrices $H_1$ and $H_2$ are equivalent but not $P$-equivalent. Let us consider the $d=3$ example given in Eq.~\eqref{eq:Hqutrit}. Here $H_1$ is in fact the discrete Fourier transform
for $d=3$, so $\ket{\psi_{G,H_1}}$
is the usual graph state.
Notice that in fact $C^{H_2}=(C^{H_1})^2$,
consequently $\ket{\psi_{G,H_2}}$
is a `weighted graph' state with edge weight
$2$ for each edge. And it is known that
the weighted graph states are in general
not LU equivalent to the `unweighted graph'
states~\cite{danielsen2008graph}.

Although in general two equivalent Hadamard matrices $H_1$ and $H_2$ may correspond to LU inequivalent generalized graph states $\ket{\psi_{G,H_1}}$ and $\ket{\psi_{G,H_2}}$,
one would ask for what kind of graphs that $\ket{\psi_{G,H_1}}$ and $\ket{\psi_{G,H_2}}$ are LU equivalent. We will show that it is the case if $G$ is a bipartite graph. That is,  the vertex set $V(G)$ of $G$ can be divided into two disjoint sets $V_1$ and $V_2$ such that there does not exists any edge $ab\in E(G)$ with $a\in V_1$ and $b\in V_2$, i.e. every edge of $G$ connects on vertex in $V_1$ to another one in $V_2$.

\begin{theorem}
If $G$ is a bipartite graph, then $\ket{\psi_{G,H_1}}$ and $\ket{\psi_{G,H_2}}$ are LU equivalent for two equivalent symmetric Hadamard matrices $H_1$ and $H_2$.
\end{theorem}
\textbf{Proof}: According to Lemma \ref{lm:phase}, we only need to deal with $H_2=P_1H_1P^T_2$, where $P_1$ and $P_2$ are two permutation matrices. Since both $H_1$ and $H_2$ are symmetric, we also have $H_2=P_2H_1P_1^T$. According to Lemma \ref{lm:P}, we then have
\begin{eqnarray}
C^{H_2}=(P_1\otimes P_2)C^{H_1}(P_1^{T}\otimes P^T_2)
=(P_2\otimes P_1)C^{H_1}(P^T_2\otimes P^T_1).
\end{eqnarray}
This means that for implementing each $C^{H_2}$ in terms of $C^{H_1}$ and single-qudit permutation operations $P_1/P_1^{T}$ and $P_2/P_2^{T}$, we can choose which of the two qudits (that $C^{H_2}$ is acting on) to apply $P_1/P_1^{T}$ or ($P_2/P_2^{T}$) on.
Notice that for bipartite graph $G$ with $V(G)=V_1\cup V_2$, we can than apply $P_1/P_1^{T}$s on vertices in $V_1$, and apply $P_2/P_2^{T}$s on vertices in $V_2$. Then the argument of \ref{th:Peq} will follow for this case, where $P_1P_1^T=P_2P_2^T=I$
$\square$.

\begin{figure}[h!]
\centerline{
\footnotesize\unitlength3\unitlength
\begin{picture}(50,35)(0,5)
\put(13,34){\circle*{3}}
\put(13,34){\line(0,-1){24}}
\put(37,34){\line(0,-1){24}}
\put(37,34){\circle*{3}}
\put(40,34){\makebox(0,0){$4$}}
\put(37,34){\line(-1,0){24}}
\put(13,34){\line(1,0){24}}
\put(10,34){\makebox(0,0){$2$}}
\put(13,10){\circle*{3}}
\put(10,10){\makebox(0,0){$1$}}
\put(37,10){\circle*{3}}
\put(40,10){\makebox(0,0){$3$}}
\put(37,10){\line(-1,0){24}}
\put(37,10){\line(0,1){24}}
\put(13,10){\line(1,0){24}}
\put(13,10){\line(0,1){24}}
\end{picture}}
\caption{An $n=4$ bipartite graph $G$, with the vertices set $V(G)=V_1\cup V_2$, where $V_1=\{1,3\}$
and $V_2=\{2,4\}$.}
\label{fig:bipartite}
\end{figure}
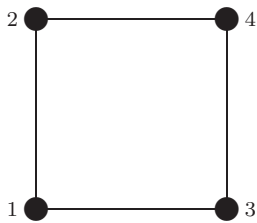

As an example, we consider a bipartite graph $G$ of $n=4$, as shown in Fig.~\ref{fig:bipartite}.

\begin{figure}[h!]

\def\gAxA{\op{H_2}\w\A{gAxA}}
\def\gAxB{\op{H_2}\w\A{gAxB}}
\def\gAxC{\op{H_2}\w\A{gAxC}}
\def\gAxD{\op{H_2}\w\A{gAxD}}
\def\gBxA{\op{P_1}\w\A{gBxA}}
\def\gBxB{\op{P_2}\w\A{gBxB}}
\def\gBxC{\op{P_1}\w\A{gBxC}}
\def\gBxD{\op{P_2}\w\A{gBxD}}
\def\gCxA{\b\w\A{gCxA}}
\def\gCxB{\b\w\A{gCxB}}
\def\gDxA{\op{P_1^T}\w\A{gDxA}}
\def\gDxB{\op{P_2^T}\w\A{gDxB}}
\def\gExB{\op{P_2}\w\A{gExB}}
\def\gFxB{\b\w\A{gFxB}}
\def\gFxC{\b\w\A{gFxC}}
\def\gGxC{\op{P_1^T}\w\A{gGxC}}
\def\gHxC{\op{P_1}\w\A{gHxC}}
\def\gIxC{\b\w\A{gIxC}}
\def\gIxD{\b\w\A{gIxD}}
\def\gJxD{\op{P_2^T}\w\A{gJxD}}
\def\gExA{*-{}\w\A{gExA}}
\def\gFxA{*-{}\w\A{gFxA}}
\def\gGxA{*-{}\w\A{gGxA}}
\def\gHxA{*-{}\w\A{gHxA}}
\def\gIxA{*-{}\w\A{gIxA}}
\def\gJxA{*-{}\w\A{gJxA}}
\def\gKxA{\op{P_1}\w\A{gKxA}}
\def\gKxD{\op{P_2}\w\A{gKxD}}
\def\gLxA{\b\w\A{gLxA}}
\def\gLxD{\b\w\A{gLxD}}
\def\gMxA{\op{P_1^T}\w\A{gMxA}}
\def\gMxD{\op{P_2^T}\w\A{gMxD}}
\def\gGxB{*-{}\w\A{gGxB}}
\def\gHxB{*-{}\w\A{gHxB}}
\def\gIxB{*-{}\w\A{gIxB}}
\def\gJxB{*-{}\w\A{gJxB}}
\def\gKxB{*-{}\w\A{gKxB}}
\def\gLxB{*-{}\w\A{gLxB}}
\def\gMxB{\op{P_2^T}\w\A{gMxB}}
\def\gJxC{*-{}\w\A{gJxC}}
\def\gKxC{*-{}\w\A{gKxC}}
\def\gLxC{*-{}\w\A{gLxC}}
\def\gMxC{\op{P_1^T}\w\A{gMxC}}

\def\bA{\ket{0}}
\def\bB{\ket{0}}
\def\bC{\ket{0}}
\def\bD{\ket{0}}

~~~~~~~~~~
\xymatrix@R=5pt@C=10pt{
    \bA & \gAxA &\gBxA &\gCxA &\gDxA &\gExA &\gFxA &\gGxA &\gHxA &\gIxA &\gJxA &\gKxA &\gLxA &\gMxA &\n
\\  \bB & \gAxB &\gBxB &\gCxB &\gDxB &\gExB &\gFxB &\gGxB &\gHxB &\gIxB &\gJxB &\gKxB &\gLxB &\gMxB &\n
\\  \bC & \gAxC &\gBxC &\n   &\n   &\n   &\gFxC &\gGxC &\gHxC &\gIxC &\gJxC &\gKxC &\gLxC &\gMxC &\n
\\  \bD & \gAxD &\gBxD &\n   &\n   &\n   &\n   &\n   &\n   &\gIxD &\gJxD &\gKxD &\gLxD &\gMxD &\n
\ar@{-}"gCxA";"gCxB"
\ar@{-}"gFxB";"gFxC"
\ar@{-}"gIxC";"gIxD"
\ar@{-}"gLxA";"gLxD"
}
\caption{The LU equivalence of two generalized graph states $\ket{\psi_{G,H_1}}$ and $\ket{\psi_{G,H_2}}$, where $G$ is the bipartite graph as shown in Fig.~\ref{fig:bipartite}. The two Hadamard matrices $H_1$ and $H_2$ satisfy $H_2=P_1H_1P_2^T$ for some permeation matrices $P_1$ and $P_2$. Two black diamonds connected by a line represents $C^{H_1}$. This circuit generates the state $\ket{\psi_{G,H_2}}$, which is given by Lemmas~\ref{lm:phase} and \ref{lm:P},
from the circuit generating  of $\ket{\psi_{G,H_1}}$. Notice that each $P_1P_1^T=P_2P_2^T=I$, so they do cancel. And $P_1H_1P_2^T\ket{0}=DH_1\ket{0}$ for some diagonal matrix $D$, which commutes with all the diagonal $C^{H_1}$s. This then shows that $\ket{\psi_{G,H_2}}=(P_1^TD)\otimes
(P_2^TD)\otimes(P_1^TD)\otimes
(P_2^TD)\ket{\psi_{G,H_1}}$.}
\label{fig:eqPbi}
\end{figure}
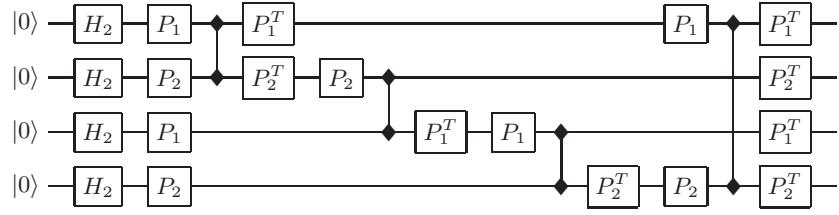

Now consider two Hadamard matrices $H_2=P_1H_1P_2^T$. The LU equivalence of the corresponding
generalized graph states $\ket{\psi_{G,H_1}}$ and $\ket{\psi_{G,H_2}}$ is then shown in Fig.~\ref{fig:eqPbi}.

\subsection{Local symmetries}

Due to Theorem \ref{th:Peq}, we will assume the Hadamard matrix $H$ has entries $1$ in the first row and first column in the following discussion.

\begin{definition}
For a symmetric Hadamard matrix $H$, if there is a
$d\times d$ permutation $P$ and a $d\times d$ diagonal unitary $D$ such that
\begin{equation}
PHD=H,
\end{equation}
then the pair $(P,D)$ is called a $S$-symmetry of $H$.
\end{definition}

\begin{proposition}
Let $G$ be any graph with $n$ vertices, and let $H$ be a $d \times d$ Hadamard. For any $S$-symmetry, $(P,D)$, the graph state $\ket{\psi_{G,H}}$ is stabilized by $P_a\prod\limits_{ab \in E(G)} D_b$ for any $a \in V(G)$.
\end{proposition}

$\textbf{Proof}$:   For any $a \in V(G)$, let $E_a = \{ij \in E(G)| i = a \textrm{ or } j=a  \}$, $\mathcal{A}_a = \prod\limits_{ij \in E_a} C_{ij}^{H} $, and let $\mathcal{B}_a = \prod\limits_{ij \in E(G)\setminus E_a} C_{ij}^{H} $. Then we have $\ket{\psi_{G,H}} = d^{-\frac{n}{2}}\mathcal{B}_a \mathcal{A}_a H^{\otimes n} |0\rangle^{\otimes n}$.

By Lemma \ref{lm:phase} and Lemma \ref{lm:P}, $C^H = C^{PHD} = (P \otimes D) C^H (P^{T} \otimes I)$. Then 
\begin{eqnarray}
\mathcal{A}_a = \prod\limits_{ab \in E_a} C_{ab}^{PHD} =  P_a (\prod\limits_{ab \in E_a} D_b) (\prod\limits_{ab \in E_a} C_{ab}^{H}) P_a^{T} = P_a (\prod\limits_{ab \in E_a} D_b) \mathcal{A}_a P_a^{T}.
\end{eqnarray}

Note that $P_a^T H_a |0\rangle = H_a |0\rangle$, and $D_b$ and $P_a$  both commute with $\mathcal{B}_a$. Therefore, we have
\begin{eqnarray}
\ket{\psi_{G,H}} & = & d^{-\frac{n}{2}} \mathcal{B}_a \mathcal{A}_a H^{\otimes n} |0\rangle^{\otimes n}\nonumber \\
                    & = &d^{-\frac{n}{2}}\mathcal{B}_a  P_a (\prod\limits_{ab \in E_a} D_b) \mathcal{A}_a P_a^{T} H^{\otimes n} |0\rangle^{\otimes n}\nonumber\\
                    & = &d^{-\frac{n}{2}}   P_a (\prod\limits_{ab \in E_a} D_b) \mathcal{B}_a \mathcal{A}_a  H^{\otimes n} |0\rangle^{\otimes n}\nonumber \\
                       & = &   P_a \prod\limits_{ab \in E_a} D_b \ket{\psi_{G,H}}.\quad \square
\end{eqnarray}

For example, if we take $H$ to be the discrete Fourier transform $\mathcal{F}_d$, and let $P, D$ be the pauli operators $X^{\dag}_d, Z_d$, respectively. Then $X^{\dag}_d \mathcal{F}_d Z_d = \mathcal{F}_d$, and the local symmetry is given by $(X^{\dag}_d)_a\prod\limits_{ab \in E(G)} (Z_d)_b$,  which is consistent with Eq.~\eqref{eq:sym}, and we recover Theorem $1$ in ~\cite{zhou2003quantum}.

Another example is the family of Hadamard matrices $H_{\alpha}$ in dimension $4$ given in Equation \ref{eq:hadamardd4}. Let $P, D$ be given as follows.
\begin{equation}
P =
\begin{pmatrix}
0 & 1 & 0 & 0 \\
1 & 0 & 0 & 0 \\
0 & 0 & 0 & 1 \\
0 & 0 & 1 & 0 \\
\end{pmatrix},
\qquad \qquad
D =
\begin{pmatrix}
1 & 0 & 0 & 0 \\
0 & 1 & 0 & 0 \\
0 & 0 & -1 & 0 \\
0 & 0 & 0 & -1 \\
\end{pmatrix}.
\end{equation}

Then $P H_{\alpha} D = H_{\alpha}.$ Notice that $P(=P^{\dag})$ and $D$ commute, so they cannot be Pauli
$X_d$ and $Z_d$ operators on the same qudit. In this sense we provide a natural generalization
of the Pauli $(X_d,Z_d)$ pairs.

\section{The triangle graph}

As a concrete example to discuss the different between a generalized graph state and a usual graph state, we use the triangle graph. As given by Theorem~\ref{th:Peq}, we only need to discuss
the case where the Hadamard matrix $H$ is `dephased', that is, the matrix elements of the first row/column are all $1$s. This can be always achieved by $D_1HD_2$, i.e. multiplying diagonal matrices from left and right, and the resulted graph states will be just LU equivalent.

The triangle $\triangle$ has the edge set $E(G)=\{(12),(23),(13)\}$, as
given by Figure $1(c)$. This gives
\begin{equation}
\ket{\psi_{\triangle,H}}=\frac{1}{d^{3/2}}C^{H}_{12}C^{H}_{23}C^{H}_{13}\sum_{ijk}\ket{ijk}
=\frac{1}{d^{3/2}}\sum_{ijk}h_{ij}h_{jk}h_{ik}\ket{ijk}.
\end{equation}

The structure of $\ket{\psi_{\triangle,H}}$ is less obvious. That is, we would like to
know whether these $\ket{\psi_{\triangle,H}}$ may be LU equivalent to each
other, for different choices of $H$. We start from the following lemma.

\begin{lemma}
\label{lm:FGHZ}
If $H$ is the discrete Fourier transform $\mathcal{F}_d$, then $\ket{\psi_{\triangle,\mathcal{F}_d}}$ is LU equivalent to the GHZ state $\ket{\text{GHZ}_{3,d}}$.
\end{lemma}

\textbf{Proof}: We use stabilizer formalism. Consider the generalized Pauli matrices $X_d,Z_d$ satisfying
\begin{equation}
X_dZ_d=q_dZ_dX_d,
\end{equation}
where $q_d=\exp{(i2\pi/d)}$. Then $\ket{\psi_{\triangle,\mathcal{F}_n}}$ is stabilized by
the stabilizer group generated by $g_1,g_2,g_3$, given by
\begin{eqnarray}
g_1&=& X^{\dag}_d \otimes Z_d \otimes Z_d \nonumber\\
g_2&=& Z_d \otimes X^{\dag}_d \otimes Z_d \nonumber\\
g_3&=& Z_d \otimes Z_d \otimes X^{\dag}_d
\end{eqnarray}

We now choose another set of generators
\begin{eqnarray}
g'_1&=&g_1=X^{\dag}_d \otimes Z_d \otimes Z_d\nonumber\\
g'_2&=&g_1^{\dag}g_2=(X_dZ_d)\otimes(Z^{\dag}_dX^{\dag}_d)\otimes I\nonumber\\
g'_3&=&g_2^{\dag}g_3=I \otimes (X_dZ_d)\otimes(Z^{\dag}_dX^{\dag}_d).
\end{eqnarray}
Notice that
\begin{equation}
(X_dZ_d)Z_d=q_dZ_d(X_dZ_d),
\end{equation}
there exists a local Clifford (LC) transformation which maps
\begin{equation}
(X_dZ_d)\rightarrow X_d,\ Z_d\rightarrow Z_d.
\end{equation}
Applying this transform on all of the three qudits maps
\begin{eqnarray}
g'_1&\rightarrow&g_1=Z_dX^{\dag}_d \otimes Z_d \otimes Z_d\nonumber\\
g'_2&\rightarrow&g_1^{\dag}g_2=X_d\otimes X^{\dag}_d\otimes I\nonumber\\
g'_3&\rightarrow&g_2^{\dag}g_3=I \otimes X_d\otimes X^{\dag}_d.
\end{eqnarray}
Furthermore, since
\begin{equation}
X_d(Z_dX^{\dag}_d)=q_d(Z_dX^{\dag}_d)X_d,
\end{equation}
there exists an LC transformation which maps
\begin{equation}
X_d\rightarrow X_d,\ Z_dX^{\dag}_d\rightarrow Z_d.
\end{equation}
Applying this transform on the first qudit maps
\begin{eqnarray}
g'_1&\rightarrow&g_1=Z_d \otimes Z_d \otimes Z_d\nonumber\\
g'_2&\rightarrow&g_1^{\dag}g_2=X_d\otimes X^{\dag}_d\otimes I\nonumber\\
g'_3&\rightarrow&g_2^{\dag}g_3=I \otimes X_d\otimes X^{\dag}_d.
\end{eqnarray}
which is LU equivalent to the GHZ state $\ket{\text{GHZ}_{3,d}}$ $\square$.

When $H$ is not the discrete Fourier transform $\mathcal{F}_n$,
$\ket{\psi_{\triangle,\mathcal{F}_n}}$ may still be
LU equivalent to $\ket{\text{GHZ}_{3,d}}$.
In fact, this is true for all $d=2,3,4,5$, which can be shown based
on the classification of complex Hadamard matrices in these
dimensions, as given by the following theorem.

\begin{theorem}
For $d=2,3,4,5$ $\ket{\psi_{\triangle,H}}$ is LU equivalent to the GHZ state $\ket{\text{GHZ}_{3,d}}$
 for any $H$.
\end{theorem}

\textbf{Proof}: We first prove this theorem up to equivalence of Hadamard matrices.
For $d=2,3,5$, any complex Hadamard matrix are equivalent to the discrete Fourier transform $\mathcal{F}_d$, which is then covered by Lemma~\ref{lm:FGHZ}.
And for $d=4$ and $H_{\alpha}$, $\ket{\psi_{\triangle,H_{\alpha}}}$ is a GHZ state as follows,
\begin{eqnarray}
\ket{\psi_{\triangle,H_{\alpha}}}&=&(\ket{0}+\ket{1}+e^{i\alpha/2}\ket{2}
-e^{i\alpha/2}\ket{3})^{\otimes 3}\nonumber\\
&+&(\ket{0}+\ket{1}-e^{i\alpha/2}\ket{2}+e^{i\alpha/2}\ket{3})^{\otimes 3}\nonumber\\
&+&e^{-3i\alpha/2}[(-\ket{0}+\ket{1}+e^{3i\alpha/2}\ket{2}+e^{3i\alpha/2}\ket{3})^{\otimes 3}\nonumber\\
&+&(\ket{0}-\ket{1}+e^{3i\alpha/2}\ket{2}+e^{3i\alpha/2}\ket{3})^{\otimes 3}].
\end{eqnarray}

Now for $d=2,3,5$, we need to deal with the cases where
$PD\mathcal{F}_d$ are still symmetric, for some permutation matrix $P$
and some diagonal matrix $D$.

For $d=2$, such $P$ and $D$ have to be identity.

For $d=3$, such $P,D$ can be identity or $PD\mathcal{F}_3=\mathcal{F}_3^*$, the later one of course generates a GHZ state for triangle graph.

For $d=5$, such $P,D$ has to satisfy that $$PD\mathcal{F}_5=\mathcal{F}_5(w^k)=(w^{ijk})_{5\times 5}, 1\leq k\leq 4$$
employing the above lemma, such Hadamard matrix generates a GHZ state for triangle graph by using $w^{k}$ instead of $w=e^{2\pi i/5}$.

For $d=4$, we also need to deal with the cases where $PDH_{\alpha}$ are still symmetric, for some permutation matrix $P$
and some diagonal matrix $D$.

Case 1, $e^{i\alpha}\neq \pm 1$,
such $P,D$ has to satisfy that
\begin{equation}
PDH_{\alpha}=
\begin{pmatrix}
1&1&1&1\\ 1&1&-1&-1\\1&-1&e^{i\alpha}&-e^{i\alpha}\\1&-1&-e^{i\alpha}&e^{i\alpha}
\end{pmatrix} or
\begin{pmatrix}
1&1&1&1\\ 1&1&-1&-1\\1&-1&-e^{i\alpha}&e^{i\alpha}\\1&-1&e^{i\alpha}&-e^{i\alpha}
\end{pmatrix}
\end{equation}
such Hadamard matrix generates a GHZ state for triangle graph by using $e^{i\alpha}$ instead of $-e^{i\alpha}$.

Case 2, $e^{i\alpha}=\pm 1$,
such $P,D$ has to satisfy that $PDH_{\alpha}$ equals on of the following matrices
\begin{equation}
\tilde{H}_a=\begin{pmatrix}
1&1&1&1\\ 1&1&-1&-1\\1&-1&1&-1\\1&-1&-1&1
\end{pmatrix},\
\tilde{H}_b=\begin{pmatrix}
1&1&1&1\\ 1&1&-1&-1\\1&-1&-1&1\\1&-1&1&-1
\end{pmatrix},\
\tilde{H}_c=\begin{pmatrix}
1&1&1&1\\ 1&-1&1&-1\\1&1&-1&-1\\1&-1&-1&1
\end{pmatrix},\
\tilde{H}_d=\begin{pmatrix}
1&1&1&1\\ 1&-1&-1&1\\1&-1&1&-1\\1&1&-1&-1
\end{pmatrix}.
\end{equation}

The fact that $\tilde{H}_a$ ($\tilde{H}_b$) corresponds to a generalized graph states
$\ket{\psi_{\triangle,\tilde{H}_a}}$ ($\ket{\psi_{\triangle,\tilde{H}_a}}$)
that is LU equivalent to $\ket{\text{GHZ}_{3,d}}$
simply follows from the previous argument for general $\alpha$ for $H_{\alpha}$.

The third matrix $\tilde{H}_c=\begin{pmatrix}
1&1\\1& -1
\end{pmatrix}\otimes \begin{pmatrix}
1&1\\1& -1
\end{pmatrix}$, so it corresponds to a generalized graph state
$\ket{\psi_{\triangle,\tilde{H}_c}}$
that is LU equivalent to $\ket{\text{GHZ}_{3,d}}$, using Proposition~\ref{pro:tensor}.

For the last matrix $\tilde{H}_d$, we notice that
\begin{equation}
\tilde{H}_d
=
\text{CNOT}\tilde{H}_c\text{CNOT},
\end{equation}
where
\begin{equation}
\text{CNOT}=\begin{pmatrix}
1&0&0&0\\ 0&1&0&0\\0&0&0&1\\0&0&1&0
\end{pmatrix}
\end{equation}
is just the usual controlled-NOT operation on two-qubits.
Then it directly follows from Theorem \ref{th:Peq} that the corresponding generalized graph state
$\ket{\psi_{\triangle,\tilde{H}_d}}$
is LU equivalent to $\ket{\text{GHZ}_{3,d}}$, for the triangle graph. We give the circuit explicitly in
Fig.~\ref{fig:case4}
$\square$.

\begin{figure}

\def\gAxA{\r\w\A{gAxA}}
\def\gAxB{\o\w\A{gAxB}}
\def\gAxC{\r\w\A{gAxC}}
\def\gAxD{\o\w\A{gAxD}}
\def\gAxE{\r\w\A{gAxE}}
\def\gAxF{\o\w\A{gAxF}}
\def\gBxA{\op{H}\w\A{gBxA}}
\def\gBxB{\op{H}\w\A{gBxB}}
\def\gBxC{\op{H}\w\A{gBxC}}
\def\gBxD{\op{H}\w\A{gBxD}}
\def\gBxE{\op{H}\w\A{gBxE}}
\def\gBxF{\op{H}\w\A{gBxF}}
\def\gCxA{\r\w\A{gCxB}}
\def\gCxB{\o\w\A{gCxA}}
\def\gCxC{\r\w\A{gCxC}}
\def\gCxD{\o\w\A{gCxD}}
\def\gCxE{\r\w\A{gCxE}}
\def\gCxF{\o\w\A{gCxF}}
\def\gDxA{\r\w\A{gDxA}}
\def\gDxB{\o\w\A{gDxB}}
\def\gDxC{\r\w\A{gDxC}}
\def\gDxD{\o\w\A{gDxD}}
\def\gDxE{\r\w\A{gDxE}}
\def\gDxF{\o\w\A{gDxF}}
\def\gExA{\r\w\A{gExA}}
\def\gExC{\r\w\A{gExC}}
\def\gExB{*-{}\w\A{gExB}}
\def\gExD{*-{}\w\A{gExD}}
\def\gFxB{\r\w\A{gFxB}}
\def\gFxD{\r\w\A{gFxD}}
\def\gGxA{\r\w\A{gGxA}}
\def\gGxB{\o\w\A{gGxB}}
\def\gGxC{\r\w\A{gGxC}}
\def\gGxD{\o\w\A{gGxD}}
\def\gHxA{\r\w\A{gHxA}}
\def\gHxB{\o\w\A{gHxB}}
\def\gHxC{\r\w\A{gHxC}}
\def\gHxD{\o\w\A{gHxD}}
\def\gIxC{*-{}\w\A{gIxC}}
\def\gExE{*-{}\w\A{gExE}}
\def\gJxC{\r\w\A{gJxC}}
\def\gJxE{\r\w\A{gJxE}}
\def\gIxD{*-{}\w\A{gIxD}}
\def\gExF{*-{}\w\A{gExF}}
\def\gJxD{*-{}\w\A{gJxD}}
\def\gKxD{\r\w\A{gKxD}}
\def\gKxF{\r\w\A{gKxF}}
\def\gLxE{\r\w\A{gLxE}}
\def\gLxF{\o\w\A{gLxF}}
\def\gMxE{\r\w\A{gMxE}}
\def\gMxF{\o\w\A{gMxF}}
\def\gIxB{*-{}\w\A{gIxB}}
\def\gNxF{*-{}\w\A{gNxF}}
\def\gIxA{*-{}\w\A{gIxA}}
\def\gJxA{*-{}\w\A{gJxA}}
\def\gKxA{*-{}\w\A{gKxA}}
\def\gLxA{*-{}\w\A{gLxA}}
\def\gNxA{\r\w\A{gNxA}}
\def\gNxE{\r\w\A{gNxE}}
\def\gOxE{*-{}\w\A{gOxE}}
\def\gOxF{*-{}\w\A{gOxF}}
\def\gPxB{\r\w\A{gPxB}}
\def\gPxF{\r\w\A{gPxF}}
\def\gQxA{\r\w\A{gQxA}}
\def\gQxB{\o\w\A{gQxB}}
\def\gKxC{*-{}\w\A{gKxC}}
\def\gLxC{*-{}\w\A{gLxC}}
\def\gMxC{*-{}\w\A{gMxC}}
\def\gNxC{*-{}\w\A{gNxC}}
\def\gOxC{*-{}\w\A{gOxC}}
\def\gPxC{*-{}\w\A{gPxC}}
\def\gQxC{\r\w\A{gQxC}}
\def\gQxD{\o\w\A{gQxD}}
\def\gQxE{\r\w\A{gQxE}}
\def\gQxF{\o\w\A{gQxF}}

\def\bA{|0\rangle_{1_a}}
\def\bB{|0\rangle_{1_b}}
\def\bC{|0\rangle_{2_a}}
\def\bD{|0\rangle_{2_b}}
\def\bE{|0\rangle_{3_a}}
\def\bF{|0\rangle_{3_b}}

~~~\xymatrix@R=5pt@C=10pt{
    \bA & \gAxA &\gBxA &\gCxA &\gDxA &\gExA &\n   &\gGxA &\gHxA &\gIxA &\gJxA &\gKxA &\gLxA &\n   &\gNxA &\n   &\n   &\gQxA &\n
\\  \bB & \gAxB &\gBxB &\gCxB &\gDxB &\gExB &\gFxB &\gGxB &\gHxB &\gIxB &\n   &\n   &\n   &\n   &\n   &\n   &\gPxB &\gQxB &\n
\\  \bC & \gAxC &\gBxC &\gCxC &\gDxC &\gExC &\n   &\gGxC &\gHxC &\gIxC &\gJxC &\gKxC &\gLxC &\gMxC &\gNxC &\gOxC &\gPxC &\gQxC &\n
\\  \bD & \gAxD &\gBxD &\gCxD &\gDxD &\gExD &\gFxD &\gGxD &\gHxD &\gIxD &\gJxD &\gKxD &\n   &\n   &\n   &\n   &\n   &\gQxD &\n
\\  \bE & \gAxE &\gBxE &\gCxE &\gDxE &\gExE &\n   &\n   &\n   &\n   &\gJxE &\n   &\gLxE &\gMxE &\gNxE &\gOxE &\n   &\gQxE &\n
\\  \bF & \gAxF &\gBxF &\gCxF &\gDxF &\gExF &\n   &\n   &\n   &\n   &\n   &\gKxF &\gLxF &\gMxF &\gNxF &\gOxF &\gPxF &\gQxF &\n
\ar@{-}"gAxB";"gAxA"
\ar@{-}"gAxD";"gAxC"
\ar@{-}"gAxF";"gAxE"
\ar@{-}"gCxB";"gCxA"
\ar@{-}"gCxD";"gCxC"
\ar@{-}"gCxF";"gCxE"
\ar@{-}"gDxB";"gDxA"
\ar@{-}"gDxD";"gDxC"
\ar@{-}"gDxF";"gDxE"
\ar@{-}"gExA";"gExC"
\ar@{-}"gFxB";"gFxD"
\ar@{-}"gGxB";"gGxA"
\ar@{-}"gGxD";"gGxC"
\ar@{-}"gHxB";"gHxA"
\ar@{-}"gHxD";"gHxC"
\ar@{-}"gJxC";"gJxE"
\ar@{-}"gKxD";"gKxF"
\ar@{-}"gLxF";"gLxE"
\ar@{-}"gMxF";"gMxE"
\ar@{-}"gNxA";"gNxE"
\ar@{-}"gPxB";"gPxF"
\ar@{-}"gQxB";"gQxA"
\ar@{-}"gQxD";"gQxC"
\ar@{-}"gQxF";"gQxE"
}
\caption{$\ket{i_a}\ket{i_b}$
for $i=1,2,3$ is the input $\ket{0}$ state for the $i$th qudit. $H$ is the single-qubit Hadamard transform (multiply by a factor of $1/\sqrt{2}$).
The two Hadamard matrices $\tilde{H}_c$ and $\tilde{H}_d$ satisfy $\tilde{H}_d=\text{CNOT}\tilde{H}_c\text{CNOT}$. Two black dots connected by a line represents the usual controlled-Z operation. One black dot connected with an $\oplus$ by a line represents the usual controlled-NOT operation, with the black dot denotes the controlled qubit. This circuit generates the state $\ket{\psi_{\triangle,\tilde{H}_d}}$, which is given by Lemmas~\ref{lm:phase} and \ref{lm:P},
from the circuit generating  of $\ket{\psi_{\triangle,\tilde{H}_d}}$. Notice that $\text{CNOT}^2=I$, so two \text{CNOT}s do cancel. And $\text{CNOT}\ket{0}_{i_a}\ket{0}_{i_b}=\ket{0}_{i_a}\ket{0}_{i_b}$. This then shows that $\ket{\psi_{\triangle,\tilde{H}_d}}=(\text{CNOT})^{\otimes 3}\ket{\psi_{\triangle,\tilde{H}_c}}$.}
\label{fig:case4}
\end{figure}

The case for $d\geq 6$ is much more complicated, as we know
there is no classification of Hadamard matrices in these higher dimensions.
However, we can indeed show that for $d=6$, some of the choices
of $H$ give the states
$\ket{\psi_{\triangle,H}}$ which are not GHZ states.
This shows that even for a small $n=3$, the generalized graph states
$\ket{\psi_{G,H}}$ may not be local unitary equivalent
to a usual graph state.

\begin{proposition}
\label{pro:nGHZ}
In general, $\ket{\psi_{\triangle,H}}$ may not be LU equivalent to $\ket{\text{GHZ}_{3,d}}$.
\end{proposition}

\textbf{Proof}: We use some known results based on invariant theory. We consider the degree $6$ invariants as
discused in~\cite{szalay2012all}. The LU invariant we compute is
\begin{equation}
I_6=\tr(\rho_{12}^{T_1})^3,
\end{equation}
where $\rho_{12}$ is the reduced density matrix (RDM)
of the $1,2$ qudits, and $T_1$ is the partial transpose on qubit $1$.

For the $\ket{\text{GHZ}_{3,6}}$, we have
\begin{equation}
I_6(\ket{\text{GHZ}_{3,6}})=0.0278.
\end{equation}

Now we consider the generalized graph state $\ket{\psi_{\triangle,H}}$ with
\begin{equation}
\label{eq:Hd6}
H=
\begin{pmatrix}
1&1&1&1 & 1 & 1\\
1&-1&i&-i & -i & i\\
1&i&-1&i & -i & -i\\
1&-i&i&-1 & i & -i\\
1&-i&-i&i & -1 & i\\
1&i&-i&-i & i & -1
\end{pmatrix}.
\end{equation}
Direct computation gives
\begin{equation}
I_6(\ket{\psi_{\triangle,H}})=0.0150.
\end{equation}
This means that $\ket{\psi_{\triangle,H}}$ is not LU equivalent to $\ket{\text{GHZ}_{3,6}}$
$\square$.

We remark that alternatively, we can observe the following: a) two tripartite states are LU equivalence iff their 2-RDMs are LU equivalent; b) two bipartite mixed states $\rho_{AB}$ and $\sigma_{AB}$ are LU equivalent iff their corresponding quantum operations $\mathcal{E}$ and $\mathcal{F}$ are unitarily equivalent, $\mathcal{E}=\mathcal{U}\circ\mathcal{F}\circ\mathcal{V}$ for some unitary operations $U,V$, where $\rho_{AB}$ and $\sigma_{AB}$ are the Choi matrices of $\mathcal{E}$ and $\mathcal{F}$.
We know that the quantum operation corresponding to $\ket{\text{GHZ}_{n,d}}$ is $\mathcal{E}=\sum_{i=0}^{d-1} E_i\cdot E_i{\dag}$ with $E_i=\ket{i}\langle{i}|$. And for the matrix $H$ we choose, the corresponding quantum operation is $\mathcal{F}=\sum_{i=0}^{d-1} F_i\cdot F_i{\dag}$ with $F_i=\Gamma_iH\Gamma_i$.
Notice that if $\ket{\text{GHZ}_{n,d}}$ is LU equivalent to $\ket{\psi_{\triangle,H}}$, then $\mathcal{E}$ and $\mathcal{F}$ are unitarily equivalent. There then exists two unitary operations $U,V$ such that $UF_iV$ are all diagonal. As a direct consequence, $F_i^{\dag}F_j$ are all commute.
Now for the Hadamard matrix $H$ as given in Eq.~\eqref{eq:Hd6}, since $F_i^{\dag}F_j$ are not all commute, we can conlude that $\ket{\psi_{\triangle,H}}$ is not LU equivalent to $\ket{\text{GHZ}_{3,d}}$.

All these methods discussed above to prove Proposition~\ref{pro:nGHZ} can be directly used to test the LU properties of other generalized graph states, $\ket{\psi_{G,H}}$ for different choices of the Hadamard matrix $H$ and different graphs $G$.

\section{Generalized graph codes}

Eq.~\eqref{eq:GS} in fact defines an encoding circuit. That is, instead of starting from the state $\ket{0}^{\otimes n}$, one can start from any computational basis state. Note that for any $\ket{i}$, we have
\begin{equation}
H\ket{i}=\Gamma_{i}H\ket{0},
\end{equation}
where $\Gamma_{i}$ is the diagonal matrix with the diagonal elements the $i$th row/column
of $H$. And here we again assume that the elements of the first row/column of $H$ are all $1$s.

Therefore, for an $n$-qubit computational basis state $\ket{i_1i_2\ldots i_n}$, we have
\begin{equation}
\label{eq:alpha}
\mathcal{U}_{G,H}\ket{i_1i_2\ldots i_n}
=\frac{1}{d^{n/2}}\prod_{ij\in E(G)}C_{ij}^{H}H^{\otimes n}\ket{i_1i_2\ldots i_n}
=\bigotimes_{k=1}^n\Gamma_{i_k}\ket{\psi_{G,H}},
\end{equation}
which is LU equivalent to $\ket{\psi_{G,H}}$ up to some diagonal local unitary determined
by the columns of $H$.

Eq.~\eqref{eq:alpha} shows that the computational basis states $\ket{i_1i_2\ldots i_n}$ are mapped to orthogonal generalized graph states (corresponding to the same graph), by the encoding circuit  $\mathcal{U}_{G,H}$ as given in Eq.~\eqref{eq:GS}. For an $n$-dit classical string $\mathbf{c}=c_1c_2\ldots c_n$, denote the corresponding quantum computational basis state by $\ket{\mathbf{c}}$, and $\mathcal{U}_{G,H}\ket{\mathbf{c}}=\ket{\psi_{G,H}(\mathbf{c})}$. Then for any $n$-dit classical code ${C}$, the encoding circuit $\mathcal{U}_{G,H}$ gives a quantum code ${Q}_C$, whose dimension is the same as the cardinality of $C$, and spanned by an orthonormal basis $\ket{\psi_{G,H}(\mathbf{c})}$ for $\mathbf{c}\in C$. In this sense, we can say that the `codewords' of ${Q}_C$ are generalized graph states. This then gives a direct generalization of the graph codes~\cite{schlingemann2001quantum}, when $C$ is a linear code. More generally, it gives a direct generalization of the codeword stabilized (CWS) codes~\cite{cross2007codeword,chuang2009codeword,chen2008nonbinary,beigi2011graph} (where $H$ is the Fourier transform $\mathcal{F}_n$).

When the dimension of ${Q}_C$ is $1$, it is a generalized graph state, and we already know from Proposition~\ref{pro:nGHZ} that is it not LU equivalent to a CWS code. Here we give an example of $Q_c$
with dimension $>1$ with $d=4$ that is not LU equivalent to a CWS code of the same classical code $C$ and the same graph $G$.

Consider the triangle graph $\triangle$ and the $4\times 4$ Hadamard matrix $H_{\alpha}$ as given in Eq.~\eqref{eq:hadamardd4}. Now choose the classical code as the linear code
\begin{equation}
C=\{000,111,222,333\},
\end{equation}
then the corresponding quantum code $Q_C$ has length $3$ and encodes an $1$. And one can check that $Q_C$ has distance $2$, so using the coding theory notation, $Q_C$ is an $[[3,1,2]]_4$ code.

By calculating the weight enumerators of $Q_C$, we know that $Q_C$ is not LU equivalent to a CWS code for some $\alpha$. For instance, $\alpha=\pi/5$. Since $C$ is linear, the corresponding CWS code is in fact additive. This shows that $Q_C$ is not an additive code for some $\alpha$. This provides a systematic method to construct non-additive quantum codes from linear classical code.

\begin{figure}[h!]

\def\gAxA{\op{H}\w\A{gAxA}}
\def\gAxB{\op{H}\w\A{gAxB}}
\def\gAxC{\op{H}\w\A{gAxC}}
\def\gBxA{\b\w\A{gBxA}}
\def\gBxB{\b\w\A{gBxB}}
\def\gCxB{\b\w\A{gCxB}}
\def\gCxC{\b\w\A{gCxC}}
\def\gDxA{\b\w\A{gDxA}}
\def\gDxC{\b\w\A{gDxC}}
\def\gExA{\op{E}\w\A{gExA}}
\def\gDxB{*-{}\w\A{gDxB}}
\def\gExC{*-{}\w\A{gExC}}
\def\gFxA{\f\w\A{gFxA}}
\def\gFxB{\b\w\A{gFxB}}
\def\gGxB{\f\w\A{gGxB}}
\def\gGxC{\b\w\A{gGxC}}
\def\gHxA{\f\w\A{gHxA}}
\def\gHxC{\b\w\A{gHxC}}
\def\gIxA{\op{H^{\dag}}\w\A{gIxA}}
\def\gHxB{*-{}\w\A{gHxB}}
\def\gIxB{\op{H^{\dag}}\w\A{gIxB}}
\def\gIxC{\op{H^{\dag}}\w\A{gIxC}}

\def\bA{ \q{q_{1}}}
\def\bB{ \q{q_{2}}}
\def\bC{ \q{q_{3}}}

~~~~~\xymatrix@R=5pt@C=10pt{
     & \gAxA &\gBxA &\n   &\gDxA &\gExA &\gFxA &\n   &\gHxA &\gIxA &\n
\\   &\gAxB &\gBxB &\gCxB &\gDxB &\n   &\gFxB &\gGxB &\gHxB &\gIxB &\n
\\    &\gAxC &\n   &\gCxC &\gDxC &\gExC &\n   &\gGxC &\gHxC &\gIxC &\n
\ar@{-}"gBxA";"gBxB"
\ar@{-}"gCxB";"gCxC"
\ar@{-}"gDxA";"gDxC"
\ar@{-}"gFxA";"gFxB"
\ar@{-}"gGxB";"gGxC"
\ar@{-}"gHxA";"gHxC"
}
\label{fig:decode}
\caption{The effect of a single qudit error $E$ after decoding. Here the bar
on top of each $C^{H}_{ij}$ means ${(C^{H}_{ij}})^{\dag}$, i.e. its hermitian conjugate.}
\end{figure}
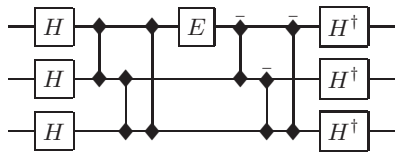

The error-correcting property of these codes would depend on both the structure of the graph $G$,
and that of the Hadamard matrix $H$. For a single-qudit error $E$, one can equivalently analyze the effect of $\mathcal{U}_{G,H} E\mathcal{U}_{G,H}^{\dag}$ on the quantum code spanned by the basis states $\ket{\mathbf{c}}$ for ${\mathbf{c}}\in C$.
For example, consider the triangle graph $\triangle$ with the encoding
circuit $\mathcal{U}_{G,H}$ as given in Fig.~\ref{fig:triangleEn}. And the circuit of $\mathcal{U}_{G,H} E\mathcal{U}_{G,H}^{\dag}$ is illustrated in Fig.~\ref{fig:decode} (for the error $E$ acting on the first qudit).

If $E$ is diagonal, then effectively on the code spanned by $\ket{\mathbf{c}}$, we still have a single qubit error, given by $HEH^{\dag}$. And, if $E$ corresponds to a generalized Pauli $X$ operator as discussed in Sec. IVB, then $HEH^{\dag}$ remains to be a tensor product of local operators, whose effect on computational basis states is relatively easy to analyze. For a general $E$, the structure of $HEH^{\dag}$ may be complicated. We will leave the analysis of the effect of errors for these generalized graph/CWS codes for future work.

\section*{Acknowledgement}

NY and BZ are supported by NSERC.

\end{document}